\documentclass[a4paper,11pt]{article}
\usepackage[compatibility=false]{caption}
\usepackage{jheppub} 
\usepackage{lineno}
\usepackage{braket}
\usepackage{cancel}
\usepackage{tikz}
\usepackage{bm}
\usepackage{braket}
\usepackage{amsmath, amssymb}
\usepackage[compat=1.1.0]{tikz-feynhand}
\usepackage[subrefformat=parens]{subcaption}
\usepackage[dvipsnames]{xcolor}
\usepackage{ulem}

\title{Feasibility Study of Lepton Number Violation in Rare $B$ and $K$ Meson Decays}

\makeatletter
\gdef\@fpheader{}
\makeatother

\preprint{KEK-TH-2791}

\author[a,b,c]{Motoi Endo,}

\author[d,a]{K\aa re Fridell,}

\author[e,f]{Sho Iwamoto,}

\author[a]{Yushi Mura,}

\author[g]{Kei Yamamoto}

\affiliation[a]{KEK Theory Center, Tsukuba, Ibaraki 305--0801, Japan}
\affiliation[b]{Graduate Institute for Advanced Studies, SOKENDAI, Tsukuba, Ibaraki 305--0801, Japan}
\affiliation[c]{Kobayashi-Maskawa Institute (KMI) for the Origin of Particles and the Universe, Nagoya University, Nagoya 464--8602, Japan}
\affiliation[d]{Institute of Particle and Nuclear Physics, Faculty of Mathematics and Physics, Charles University in Prague, V Hole\v{s}ovi\v{c}k\'{a}ch 2, 180 00 Praha 8, Czech Republic}
\affiliation[e]{Department of Physics, National Sun Yat-Sen University, Kaohsiung 80424, Taiwan}
\affiliation[f]{Center for Theoretical and Computational Physics, National Sun Yat-Sen University, Kaohsiung 80424, Taiwan}
\affiliation[g]{Faculty of Science and Engineering,
Iwate University, Morioka, Iwate 020-8550, Japan}

\abstract{
We study lepton-number-violating interactions at dimension seven in the Standard Model effective field theory that contribute to the meson decays $B \to K \nu \nu$ and $K \to \pi \nu \nu$.
Such interactions could washout the baryon asymmetry of the Universe and also contribute to the neutrinoless double beta decay, even though the interactions involve a change in down-type quark flavors.
We clarify conditions under which excesses in meson decay rates over the Standard Model predictions can be successfully observed.
We also show that, although these interactions contribute to neutrino masses at the two-loop level, the Weinberg operator can be introduced consistently without spoiling the scenario.
}

\begin{document}
\maketitle
\flushbottom

\allowdisplaybreaks

\section{Introduction \label{sec:intro}}

Evidence for non-zero neutrino masses~\cite{Super-Kamiokande:1998kpq} provides one of the most compelling hints of physics beyond the Standard Model (SM).
If neutrino masses are of Majorana type, lepton number cannot be conserved and the new physics responsible for neutrino mass must contain sources of lepton number violation (LNV), where the smallness of neutrino masses is explained by suppression with inverse powers of the new physics scale~\cite{Minkowski:1977sc,Yanagida:1979as,Gell-Mann:1979vob,Mohapatra:1979ia,Schechter:1980gr}.

At present, however, there is no direct experimental evidence to determine the scale of new physics with LNV, and such a scale could in principle lie near the TeV scale.
In this case, searches for the neutrinoless double beta ($0\nu\beta\beta$) decay provide one of the most sensitive probes of the Majorana nature of neutrinos. 
The decay rate can be probed through measurements of the isotope half-life~\cite{Vissani:1999tu,Pas:1999fc,Pas:2000vn,Deppisch:2012nb}.
From the latest KamLAND-Zen result~\cite{KamLAND-Zen:2024eml}, lower bounds on the half-life have been obtained, and significant improvements in sensitivity are expected in forthcoming experiments~\cite{Giuliani:2019uno}.

LNV interactions can also have an impact on the early Universe.
Since electroweak sphalerons violate baryon number $B$ and lepton number $\ell$\footnote{We reserve the symbol $L$ for the left-handed lepton doublet.}, whereas $B-\ell$ is conserved, a baryon asymmetry generated at high temperatures is typically tied to an initial $B-\ell$ asymmetry~\cite{Manton:1983nd,Kuzmin:1985mm,Bento:2003jv,Harvey:1990qw}.
If LNV reactions are efficient while sphalerons are active, they can erase the initial asymmetry and thereby jeopardize baryogenesis~\cite{Nelson:1990ir,Deppisch:2013jxa,Deppisch:2015yqa,Deppisch:2017ecm,Deppisch:2020oyx}.
This washout constraint has long been emphasized for LNV processes associated with the Weinberg operator~\cite{Nelson:1990ir}, and has more recently been extended to higher-dimensional LNV interactions~\cite{Deppisch:2013jxa,Deppisch:2015yqa,Deppisch:2017ecm,Deppisch:2020oyx,Fridell:2024otl}.
Consequently, any LNV interaction large enough to produce observable low-energy signals is subject to non-trivial bounds from the requirement of successful baryogenesis.

Furthermore, LNV interactions may affect semileptonic meson decays.
In particular, $B \to K \nu \overline{\nu}$ and $K \to \pi \nu \overline{\nu}$ are quark-flavor-changing neutral current processes.
In the SM, they are highly suppressed, providing sensitive probes of new physics (see, {\it e.g.}, refs.~\cite{Inami:1980fz,Buchalla:1993bv,Buchalla:1992zm,Buchalla:1993wq,Buchalla:1997kz,Buchalla:1998ba,Colangelo:1996ay,Melikhov:1998ug,Kim:1999waa,Misiak:1999yg,Aliev:2001in,Buras:2006gb,Brod:2008ss,Brod:2010hi,Altmannshofer:2011gn,Buras:2014fpa,Bause:2021cna,Li:2019fhz,Deppisch:2020oyx,Felkl:2021uxi,Fridell:2023rtr,Buras:2024ewl,Brod:2021hsj,Parrott:2022zte,Buras:2022qip,Gartner:2024muk,Fridell:2023ssf}).
Experimentally, Belle~II has recently reported evidence for a measurement of the branching fraction of $B^+ \to K^+ + \cancel{E}$, showing a $2.7\sigma$ excess over the SM prediction~\cite{Belle-II:2023esi}.
The NA62~\cite{NA62:2024pjp} and KOTO~\cite{KOTO:2024zbl} experiments have also reached impressive sensitivities in $K^+ \to \pi^+ + \cancel{E}$ and $K_L \to \pi^0 + \cancel{E}$, respectively, with further improvements expected in the near future~\cite{KOTO:2025uqg}.
Since the invisible particles responsible for missing energy $\cancel{E}$ are not identified directly in the detectors, deviations from the SM predictions may suggest contributions from the LNV interactions inducing $b\to s\nu\nu$ and $s\to d\nu\nu$ (rather than $\nu\bar\nu$) transitions~\cite{Felkl:2021uxi,Li:2019fhz,Buras:2024ewl,Fridell:2023rtr,Deppisch:2020oyx}\footnote{
Here and hereafter, the neutrino pair $\nu\nu$ represents $\nu\nu$ and $\bar\nu\bar\nu$, unless otherwise explicitly stated. 
}.

In this paper, we study the feasibility of LNV interactions affecting the meson decays. 
Within the framework of the Standard Model effective field theory (SMEFT)~\cite{Buchmuller:1985jz,Grzadkowski:2010es,Babu:2001ex,deGouvea:2007qla,Lehman:2014jma,Liao:2016hru,Brivio:2017vri,Isidori:2023pyp}, the decays $B \to K \nu \nu$ and $K \to \pi \nu \nu$ are generated by the $\overline{d}LQLH$-type operator at dim.~7.
We first revisit their implications for baryon and lepton-number washout in the early Universe.
Such effects have been studied in refs.~\cite{Deppisch:2013jxa,Deppisch:2015yqa,Deppisch:2017ecm,Deppisch:2020oyx}, where the Boltzmann equation for the evolution of the lepton asymmetry has been solved in the presence of the LNV operators. 
However, the lepton number itself is also violated by the electroweak sphaleron.
Hence, in this paper, we solve the Boltzmann equations for $B/3-\ell_\alpha$, which is affected purely by the new LNV interactions.
We show that the washout depends on the lepton-flavor structure of the LNV operators and identify a viable scenario for successful baryogenesis in their presence.
Moreover, in our framework, the LNV operators induce long-range contributions to the $0\nu\beta\beta$ decay.
Even though the operators change the down-type quark flavors, we find that they can contribute to the decay at the tree level through the Cabibbo--Kobayashi--Maskawa (CKM) matrix~\cite{Cabibbo:1963yz,Kobayashi:1973fv} or at the one-loop level via flavor-changing loop corrections on the external down-type quark legs.
We compare the current status and prospects of the rare meson decays with the $0\nu\beta\beta$ searches and the requirement that a primordial baryon asymmetry generated above the SMEFT cutoff is not excessively washed out.

We also consider neutrino masses generated by the LNV operators. 
Since the operators are assumed to change the quark flavors, these masses are induced at the two-loop level through electroweak interactions.
In this paper, we perform the two-loop calculation explicitly.
Since the contributions are found to be sizable, we consider that the experimental results for the neutrino masses are explained by additionally introducing the Weinberg operator, which violates the lepton number at dim.~5.
Then, we analyze the effects of the Weinberg operator on the baryon number washout and on the $0\nu\beta\beta$ decay.

This paper is organized as follows.
In section~\ref{sec:operator}, we introduce LNV operators at dim.~5, corresponding to the Weinberg operator, and those at dim.~7, which are relevant to our discussion.
The rare meson decays $B \to K \nu \nu$ and $K \to \pi \nu \nu$ induced by the dim.~7 LNV operators are discussed in section~\ref{sec:meson}.
In section~\ref{sec:washout}, the baryon and lepton number washout in the early Universe via those LNV operators is discussed.
In sections~\ref{sec:neutrinomass} and \ref{sec:0nuee}, we calculate the Majorana masses for the active neutrinos induced by the dim.~7 LNV operators at the two-loop level and study the $0\nu\beta\beta$ decay, respectively.
In section~\ref{sec:results}, we present the results for the cases in which the $B \to K \nu \nu$ and $K \to \pi \nu \nu$ decays are relevant.
In section~\ref{sec:discussions}, we present our discussions and conclusions.

\section{LNV operators \label{sec:operator}}
In this section, we introduce LNV operators relevant to semileptonic meson decays and their phenomenological implications.
Flavor-changing processes induced by the SM or new physics can be described by the low-energy effective field theory (LEFT)~\cite{Aebischer:2017gaw,Jenkins:2017jig,Jenkins:2017dyc}
in terms of higher dimensional operators respecting $\mathrm{U}(1)_{\mathrm{EM}}$.
If the process is affected by new physics beyond the electroweak scale, the Standard Model effective field theory (SMEFT)~\cite{Buchmuller:1985jz,Grzadkowski:2010es,Babu:2001ex,deGouvea:2007qla,Lehman:2014jma,Liao:2016hru,Brivio:2017vri,Isidori:2023pyp} can be used to encode the effect in terms of higher dimensional operators built from SM fields.
In the SMEFT, LNV interactions appear as higher-dimensional operators ($d \ge 5$) of odd mass dimensions~\cite{Babu:2001ex,deGouvea:2007qla,Lehman:2014jma,Kobach:2016ami}.

At dim.~5, the Weinberg operator is the only source of LNV, which is given by~\cite{Weinberg:1979sa},
\begin{align}
  \mathcal{L}_5 &= \sum_{\alpha\beta} \sum_{ijmn}  C_{\alpha \beta}^{\nu \nu} \epsilon_{ij} \epsilon_{mn}  (\overline{L_\beta^{i C}} L_\alpha^m) H^j H^n +\mathrm{h.c.},
  \label{eq:Weinbergop}
\end{align}
where $L$ and $H$ are the left-handed lepton doublet and the Higgs doublet, respectively.
The indices $\alpha$ and $\beta$ denote lepton flavors, and
$i,j,m$, and $n$ are $\mathrm{SU}(2)_L$ indices with $\epsilon_{ij}$ being the antisymmetric tensor ($\epsilon_{12} = +1$).
The left-handed lepton doublets are denoted as $L_\alpha = (l, \nu)_\alpha^\intercal$, where $l$ ($\nu$) denotes the left-handed charged leptons (neutrinos). 
The Wilson coefficient $C^{\nu \nu}_{\alpha \beta}$ is taken symmetric under $\alpha \leftrightarrow \beta$.

At dim.~7, the LNV operators relevant to the meson decays $B \to K \nu \nu$ and $K \to \pi \nu \nu$ are given solely by the $\overline{d}LQLH$ type~\cite{Lehman:2014jma}, 
\begin{align}
    \mathcal{L}_{7} = \sum_{pr} \sum_{\alpha \beta} \sum_{ijmn} \Big( C^{pr}_{\alpha \beta} \epsilon_{ij} \epsilon_{mn} (\overline{d_p}L_\alpha^i) (\overline{Q_r^{j C}} L^m_\beta ) H^n + \mathrm{h.c.} \Big),
    \label{eq:SMEFT_dim7}
\end{align}
where $Q$ and $d$ denote the left-handed quark doublet and the right-handed down-type quarks, respectively.
The indices $p,r$ represent the quark generations.
The left-handed quark doublets are denoted by $Q_p = (V^\dagger q_{1}, q_{2})_p^\intercal$, where $q_1$ ($q_2$) are the left-handed up- (down-) type quarks in the mass eigenstate basis, and $V$ is the CKM matrix.
The coefficient $C^{pr}_{\alpha \beta}$ is neither symmetric or antisymmetric under $\alpha \leftrightarrow \beta$.
For later analysis, we define the dimensionless couplings as
\begin{align}
    c^{pr}_{\alpha \beta} \equiv C^{pr}_{\alpha \beta} \times \Lambda^3, 
\end{align}
where $\Lambda$ is the cutoff scale in the SMEFT.

After the electroweak symmetry breaking, the LNV operators induce the following LEFT interactions~\cite{Felkl:2021uxi,Buras:2024ewl}:
\begin{align}
  \mathcal{L}_{\mathrm{LEFT}} &= \sum_{pr} \sum_{\alpha \beta} \Big( c^{\mathrm{SLL}}_{\alpha \beta p r} (\overline{\nu_{\alpha}^C} \nu_\beta) (\overline{d_p} q_{2,r}) +c^{\mathrm{TLL}}_{\alpha \beta p r}  (\overline{\nu_{\alpha}^C} \sigma^{\mu \nu} \nu_\beta) (\overline{d_p} \sigma_{\mu \nu} q_{2,r}) + \mathrm{h.c.} \Big),
  \label{eq:LEFT}
\end{align} 
where $\sigma^{\mu \nu} = \frac{i}{2} [\gamma^\mu, \gamma^\nu]$.
As shown in appendix~\ref{sec:fierz}, the LEFT coefficients are given in terms of $C^{pr}_{\alpha \beta}$ as
\begin{align}
  c_{\alpha \beta p r}^{\mathrm{SLL}} = -\frac{v}{4 \sqrt{2}} (C_{\alpha \beta}^{pr} + C_{\beta \alpha}^{pr}), ~~~ 
  c_{\alpha \beta p r}^{\mathrm{TLL}} = \frac{v}{16 \sqrt{2}} (C_{\alpha \beta}^{pr} - C_{\beta \alpha}^{pr}).
  \label{eq:matching_SMEFT_LEFT}
\end{align} 
The scalar coefficients are symmetric under $\alpha \leftrightarrow \beta$, while the tensor coefficients are antisymmetric.
It is noticed that, although the SMEFT operator \eqref{eq:SMEFT_dim7} generates both the scalar and tensor operators in the LEFT \eqref{eq:LEFT}, the latter contribution is 4 times smaller than the former.

\section{Rare meson decays \label{sec:meson}}
In this section, we discuss the rare meson decays induced by the effective operators.

\subsection{The \texorpdfstring{$B^+ \to K^+ \nu \nu$}{B+ to K+ nu nu} decay}
The measurements of $B^+\to K^++\cancel{E}$ may have contributions both from $B^+ \to K^+ \nu \overline{\nu}$ and $B^+ \to K^+ \nu \nu$~\cite{Felkl:2021uxi,Li:2019fhz,Buras:2024ewl,Fridell:2023rtr,Deppisch:2020oyx}.
The former decay proceeds in the SM, which is suppressed by the $W$ boson loops in the penguin and box diagrams.
Including long-distance contributions via the double-charged current at the tree level, the branching ratio predicted by the SM is evaluated as~\cite{Parrott:2022zte} 
\begin{align}
  \mathrm{Br}(B^+ \to K^+ \nu \overline{\nu})_{\mathrm{SM}} = (5.58 \pm 0.37) \times 10^{-6}.
\end{align}
Recently, Belle~II has reported a measurement of its branching ratio as~\cite{Belle-II:2023esi}
\begin{align}
  \mathrm{Br}(B^+ \to K^+ \nu \overline{\nu})_{\mathrm{exp}} = \big(2.3 \pm 0.5(\mathrm{stat})^{+0.5}_{-0.4} (\mathrm{syst}) \big) \times 10^{-5},
  \label{eq:BKnn_BelleII}
\end{align}
where the results of the inclusive and hadronic tagging analyses are combined.
This shows a 2.7$\sigma$ departure from the SM prediction~\cite{Belle-II:2023esi}.
At Belle~II with $50$~$\mathrm{ab}^{-1}$ in the future, an $11\%$ precision for the measurement of the branching ratio is expected~\cite{Belle-II:2018jsg}.

Since the neutrinos are not identified in the detectors, the decay $B^+ \to K^+ \nu \nu$ induced by the LNV interactions contributes to the measurements.
In eq.~\eqref{eq:LEFT}, the relevant combinations of the quark-flavor indices for this decay are $(p,r) = (2,3)$ and $(3,2)$\footnote{
In this paper, $(p,r) = (s,b)$ and $(b,s)$ are also used just for readability. 
}. 
Namely, the corresponding dim.~7 couplings are $C^{sb}_{\alpha \beta}$, which contributes to $B^+ \to K^+ \nu_\alpha \nu_\beta$, and $C^{bs}_{\alpha \beta}$, inducing $B^+ \to K^+ \bar\nu_\alpha \bar\nu_\beta$.
The LNV contribution to the branching fraction is given in terms of the LEFT coefficients as~\cite{Felkl:2021uxi,Buras:2024ewl}
\begin{equation}
    \mathrm{Br}(B^+ \to K^+ \nu \nu)_{\mathrm{NP}} = J_S^{B^+} \sum_{\alpha \le \beta} \frac{1}{1+\delta_{\alpha \beta}} \Big( |c^{\mathrm{SLL}}_{sb \alpha \beta}|^2 + |c^{\mathrm{SLL}}_{bs \alpha \beta}|^2 \Big) + J_T^{B^+} \sum_{\alpha < \beta} \Big( |c^{\mathrm{TLL}}_{sb \alpha \beta}|^2 + |c^{\mathrm{TLL}}_{bs \alpha \beta}|^2 \Big),
\end{equation}
where all possible neutrino pairs are taken into account.
The analytic expressions for the $J$ factors are given in appendix~\ref{sec:formfactor}.
In the numerical analysis, we use the following values as given in ref.~\cite{Buras:2024ewl}:
\begin{align}
    J_S^{B^+} = 10.4 G_F^{-2}, ~~~~ J_T^{B^+} = 17.7 G_F^{-2},
\end{align}
where $G_F = 1/(\sqrt{2}v^2)$ and $v\approx246~\mathrm{GeV}$.
Although the tensor $J$ factor is larger by about 1.7 than the scalar one, the Wilson coefficient of the tensor is 4 times smaller than that of the former, as mentioned in the previous section.

Recently, the Belle II collaboration has reported the reinterpretation of eq.~\eqref{eq:BKnn_BelleII} using model-agnostic likelihoods~\cite{Gartner:2024muk} and provided the allowed regions for the coefficients of the scalar and tensor operators as well as the vector ones~\cite{Belle-II:2025lfq}.
With a uniformly flat prior, $|C^{\mathrm{SL}} + C^{\mathrm{SR}}| = [0,15.4]$ and $|C^{\mathrm{TL}}| = [0,11.2]$ are obtained as the $95\%$~highest density intervals in their operator basis.
They are converted to our LEFT basis as (cf. ref.~\cite{Gartner:2024muk})
\begin{align}
  |C^{\mathrm{SL}} + C^{\mathrm{SR}}|^2 &\to \frac{2}{3} | \mathcal{A} |^{-2} \sum_{\alpha \le \beta} \frac{1}{1 + \delta_{\alpha \beta}} \Big( |c^{\mathrm{SLL}}_{\alpha \beta sb}|^2 +  |c^{\mathrm{SLL}}_{\alpha \beta bs}|^2 \Big), \notag \\
  |C^{\mathrm{TL}}|^2 &\to\frac{2}{3} | \mathcal{A} |^{-2} \sum_{\alpha < \beta} \Big( |c^{\mathrm{TLL}}_{\alpha \beta sb}|^2 +  |c^{\mathrm{TLL}}_{\alpha \beta bs}|^2 \Big),
\end{align}
where $\mathcal{A} = -(4 G_F/\sqrt{2}) (\alpha_{em}/2\pi) V_{32}^* V_{33}$, and $\alpha_{em}$ is the fine-structure constant.
Therefore, the corresponding intervals are obtained as 
\begin{align}
    \sqrt{\sum_{\alpha \le \beta} \frac{1}{1 + \delta_{\alpha \beta}} \Big( |c^{\mathrm{SLL}}_{\alpha \beta sb}|^2 +  |c^{\mathrm{SLL}}_{\alpha \beta bs}|^2 \Big)} &= [0,3.13 \times 10^{-8}]~\mathrm{GeV}^{-2}, \notag \\
    \sqrt{\sum_{\alpha < \beta} \Big( |c^{\mathrm{TLL}}_{\alpha \beta sb}|^2 +  |c^{\mathrm{TLL}}_{\alpha \beta bs}|^2 \Big)} &= [0,2.28 \times 10^{-8}]~\mathrm{GeV}^{-2}.
    \label{eq:BKvvbound}
\end{align}
Although the original SMEFT operator \eqref{eq:SMEFT_dim7} generates both the scalar and tensor operators in the LEFT as in eq.~\eqref{eq:matching_SMEFT_LEFT}, the latter contribution is less than about $10\%$ of the former one unless cancellations happen for the scalar contribution. 
We will simply neglect the contribution via the tensor operator in the following analysis.

\subsection{The \texorpdfstring{$K^+ \to \pi^+ \nu \nu$}{K+ to pi+ nu nu} and \texorpdfstring{$K_L \to \pi^0 \nu \nu$}{KL to pi0 nu nu} decays}

The decays $K^+ \to \pi^+ \nu \overline{\nu}$ and $K_L \to \pi^0 \nu \overline{\nu}$ are rare processes in the SM.
The theoretical predictions for these decays in the SM are given by~\cite{Buras:2022qip}
\begin{align}
  &\mathrm{Br}(K^+ \to \pi^+ \nu \overline{\nu})_{\mathrm{SM}} = (8.60 \pm 0.42) \times 10^{-11}, \notag \\
  &\mathrm{Br}(K_L \to \pi^0 \nu \overline{\nu})_{\mathrm{SM}} = (2.94 \pm 0.15) \times 10^{-11},
\end{align}
respectively.
On the other hand, NA62 has recently reported the following result for $K^+ \to \pi^+ \nu \overline{\nu}$~\cite{NA62:2024pjp}:
\begin{align}
  \mathrm{Br}(K^+ \to \pi^+ \nu \overline{\nu})_{\mathrm{exp}} = 13.0^{+3.3}_{-3.0} \times 10^{-11}.
\end{align}
The current 90\% C.L.~upper bound on the $K_L \to \pi^0 \nu \overline{\nu}$ branching ratio has been reported by KOTO as~\cite{KOTO:2024zbl}
\begin{align}
  \mathrm{Br}(K_L \to \pi^0 \nu \overline{\nu})_{\mathrm{exp}} < 2.2 \times 10^{-9}.
\end{align}
The goal of KOTO is to reach a sensitivity to branching ratios of $\mathcal{O}(10^{-10})$, and in the upgraded KOTO experiment (KOTO II)~\cite{KOTO:2025uqg}, a 40\% deviation from the SM prediction would correspond to a 90\% C.L.~indication of new physics.

Similarly to the previous subsection, the measurements of these K meson decays are affected by the LNV interactions.
The quark-flavor indices relevant to them are $(p,r) = (1,2)$ and $(2,1)$\footnote{
Again, $(p,r) = (d,s)$ and $(s,d)$ are also used in the following.
}, where the dim.~7 couplings $C^{ds}_{\alpha \beta}$ and $C^{sd}_{\alpha \beta}$ are involved.
In terms of the LEFT operators, the LNV contributions to the branching ratios of $K^+ \to \pi^+ \nu \nu$ and $K_L \to \pi^0 \nu \nu$ are given by~\cite{Li:2019fhz, Buras:2024ewl}\footnote{
In eq.~\eqref{eq:KLpi0nn_tensor}, the relative sign between $c^{\mathrm{TLL}}_{ds \alpha \beta}$ and $c^{\mathrm{TLL}}_{sd \alpha \beta}$ is negative in our result, which disagrees with eq.~(3.6) in ref.~\cite{Buras:2024ewl}. 
This is because the form factors satisfy $\bra{\pi^0} \overline{d} \sigma^{\mu \nu} s \ket{\overline{K^0}} = - \bra{\pi^0} \overline{s} \sigma^{\mu \nu} d \ket{K^0}$ as well as $\bra{\pi^0} \overline{d} \gamma^{\mu} s \ket{\overline{K^0}} = - \bra{\pi^0} \overline{s} \gamma^{\mu} d \ket{K^0}$ and $\bra{\pi^0} \overline{d} s \ket{\overline{K^0}} = + \bra{\pi^0} \overline{s} d \ket{K^0}$ (cf.~ref.~\cite{Ligeti:2016npd}). 
}
\begin{align}
    &\mathrm{Br}(K^+ \to \pi^+ \nu \nu)_{\mathrm{NP}} = J_S^{K^+} \sum_{\alpha \le \beta} \frac{1}{1+\delta_{\alpha \beta}} \Big( |c^{\mathrm{SLL}}_{ds \alpha \beta}|^2 + |c^{\mathrm{SLL}}_{sd \alpha \beta}|^2 \Big) + J_T^{K^+} \sum_{\alpha < \beta} \Big( |c^{\mathrm{TLL}}_{ds \alpha \beta}|^2 + |c^{\mathrm{TLL}}_{sd \alpha \beta}|^2 \Big), \\
    &\mathrm{Br}(K_L \to \pi^0 \nu \nu)_{\mathrm{NP}} = J_S^{K_L} \sum_{\alpha \le \beta} \frac{1}{1+\delta_{\alpha \beta}} |c^{\mathrm{SLL}}_{ds \alpha \beta} + c^{\mathrm{SLL}}_{sd \alpha \beta}|^2 + J_T^{K_L} \sum_{\alpha < \beta} |c^{\mathrm{TLL}}_{ds \alpha \beta} - c^{\mathrm{TLL}}_{sd \alpha \beta}|^2,
    \label{eq:KLpi0nn_tensor}
\end{align}
where the $J$ factors are given in appendix~\ref{sec:formfactor}, and we use the following values in the numerical analysis~\cite{Buras:2024ewl}:
\begin{align}
    J_S^{K^+} = 15.2 G_F^{-2}, ~~~~ J_T^{K^+} = 0.11 G_F^{-2}, ~~~~ J_S^{K_L} = 34.6 G_F^{-2}, ~~~~ J_T^{K_L} = 0.13 G_F^{-2}.
\end{align}
In contrast to $B^+ \to K^+ \nu \nu$, the tensor $J$ factors are much smaller than the scalar ones, and thus, the tensor contributions are much smaller than those from the scalar operator unless the latter is suppressed. 

\section{Baryon and lepton number washout in the early Universe \label{sec:washout}}
Let us move on to the discussion of the baryon number washout induced by the LNV operators.
At temperatures above the scale of the electroweak symmetry breaking, the electroweak sphaleron processes are active, which violate the baryon number $B$ and the lepton number $\ell$ while keeping $B-\ell$ conserved~\cite{Kuzmin:1985mm,Bento:2003jv}.
Hence, LNV interactions in this period work to washout baryon asymmetry and, if sizable, spoil typical baryogenesis scenarios~\cite{Nelson:1990ir,Deppisch:2013jxa,Deppisch:2015yqa,Deppisch:2017ecm,Deppisch:2020oyx}.

We assume the baryon asymmetry is initially generated at a higher-energy scale by some baryogenesis mechanism.
Including the LNV operators \eqref{eq:Weinbergop} and \eqref{eq:SMEFT_dim7}, we analyze the time evolution of the baryon asymmetry, observe the washout effect caused by the operators, and find constraints on their strength by requiring that the baryon asymmetry is not excessively washed out.
Although the time evolution also depends on the UV theory above the SMEFT cutoff scale, we do not include those model-dependent effects here, for which the constraints presented in this section are conservative.

\subsection{Quantum charges preserved by the SM processes}
The time evolution of the number density $n_\psi$ of a particle $\psi$ in the early Universe is described by the Boltzmann equation (cf.~refs.~\cite{Griest:1990kh, Edsjo:1997bg, Giudice:2003jh})
\begin{align}
  zH n_\gamma \frac{d \eta_{\Delta \psi}}{d z} = \mathcal{C}
\end{align}
with respect to $z = \Lambda / T$, where $T$ is the photon temperature,
$\Lambda$ is an arbitrary scale, but to be set as the SMEFT cut-off scale,
$n_\gamma = 2 \zeta(3)T^3/\pi^2$ is the photon number density,
$H \simeq 1.66 \sqrt{g_*} T^2 / \Lambda_{P}$ is the Hubble parameter with the Planck scale $\Lambda_P = 1.2 \times 10^{19}~\mathrm{GeV}$ and the effective degrees of freedom $g_* = 106.75$,
and $\eta_{\Delta\psi} = (n_\psi-n_{\overline\psi}) / n_\gamma$ is the $\psi$-asymmetry normalized by the photon density.
The right-hand side, $\mathcal C$, is the collision term discussed later, which receives contributions only from processes that change the asymmetry $\Delta\psi$.

Due to the gauge interactions in the SM, all relevant SM particles are in kinetic equilibrium in the early Universe, and the asymmetry for a particle $\psi$ is approximately given in terms of its chemical potential $\mu_\psi$ as
\begin{equation}
     \eta_{\Delta \psi} \approx \bigg( \frac{3}{4} \bigg)^F \times \frac{g_\psi}{2} \times \frac{2 \mu_{\psi}}{T},
\end{equation}
where $F=0$ ($F=1$) for bosons (fermions) and $g_\psi$ is the number of degrees of freedom.

Since the gauge interactions are in chemical equilibrium in the early Universe, the chemical potentials of the SM particles are given by the following:
$\mu_{Q}$, $\mu_{u}$, $\mu_{d}$, $\mu_{L_\alpha}$, $\mu_{e_\alpha}$, and $\mu_H$, where $\alpha$ denotes the lepton flavor index. Quark flavors are no longer respected because of CKM mixing.
If, in addition, the SM Yukawa interactions and the electroweak sphaleron processes are in chemical equilibrium, the chemical potentials are related by\footnote{
Precisely speaking, the electron Yukawa interaction is in thermal equilibrium below $T = \mathcal{O}(10^5)~\mathrm{GeV}$~\cite{Bodeker:2019ajh,Domcke:2020quw,Domcke:2020kcp}, where the right-handed electron asymmetry starts to be transferred.}
\begin{align}
  &\mu_{Q} = -\frac{1}{9} \sum_{\alpha} \mu_{L_\alpha},
 &&\mu_{u} = \frac{5}{63} \sum_{\alpha} \mu_{L_\alpha},
 &&\mu_{d} = -\frac{19}{63} \sum_{\alpha} \mu_{L_\alpha},
\notag\\
  &\mu_{H} = \frac{4}{21} \sum_\alpha \mu_{L_\alpha},
 &&\mu_{e_\alpha} = \mu_{L_\alpha} - \frac{4}{21} \sum_\beta\mu_{L_\beta}.
  \label{eq:chemical_1}
\end{align}
Three chemical potentials $\mu_{L_\alpha}$ remain unmixed, which reflects the fact that the quantum numbers $B/3-\ell_\alpha$ are preserved as long as we consider the SM processes with the electroweak sphalerons, while $B$ and $\ell_\alpha$ are not individually.
Furthermore, because of this fact, the collision term for the time evolution of $\eta_{\Delta(B/3 - \ell_{\alpha })}$ remains zero as long as we consider only SM processes.
Therefore, it is convenient to describe the time evolution by
\begin{align}
&  zH n_\gamma \frac{d \eta_{\Delta(B/3 - \ell_{\alpha })}}{d z} = \mathcal{C},
&& \eta_{\Delta(B/3-\ell_\alpha)}
  =\frac{3}{4T}\Bigl(
    - 3\mu_{L_\alpha}-\frac{16}{63} \sum_\beta \mu_{L_\beta} \Bigr).
  \label{eq:chemical_2}
\end{align}
The collision term is schematically given by
\begin{equation}
    \mathcal{C} =- \frac{2}{T} \sum_i \bigg( \sum_{\mathrm{ini}}(\mu) - \sum_{\mathrm{fin}}(\mu) \bigg) \gamma^{eq}_i
  \label{eq:Boltzmann1}
\end{equation}
where the summation is taken over all processes $i$ that change $B/3 - \ell_{\alpha}$, {\it i.e.}, the processes from the LNV operators in our case.
The summations $\sum_{\mathrm{ini}} (\star)$ ($\sum_{\mathrm{fin}} (\star)$) are taken over all initial (final) particles in a relevant process $i$.
The thermal average of the squared amplitude for the process $i$ is given by (cf.~ref.~\cite{Deppisch:2017ecm})
\begin{align}
  \gamma^{eq}_i = \int [d \Pi]_{\mathrm{ini}} [f^{eq}]_{\mathrm{ini}} \int [d \Pi]_{\mathrm{fin}} (2 \pi)^4 \delta^4 \Big(\sum_{\mathrm{ini}} (p) - \sum_{\mathrm{fin}} (p) \Big) |\mathcal{M}_i|^2,
\end{align}
where we have used a shorthand notation $[\star]_{\mathrm{ini}}$ ($[\star]_{\mathrm{fin}}$) as shorthand for the products of the phase space integral $d\Pi = g \frac{d^3 p}{(2 \pi)^3(2E)}$ and the Maxwell-Boltzmann distribution $f^{eq} = e^{-E/T}$ for all initial (final) particles in the process $i$.
We note that the inverse and CP-conjugated processes for $i$ are taken into account in eq.~\eqref{eq:Boltzmann1}.

\subsection{Washout by the LNV operators}

\begin{table}[t]
  \centering
    \begin{center}
      \begin{tabular}{|c|c|}
        \hline
        Name: $i$ & Process  \\ \hline
        ($\nu$1) & $L_{\alpha } L_{\beta } \to \overline{H}\overline{H}$  
        \\ \hline
        ($\nu$2) & $L_{\alpha } H \to \overline{L}_{\beta } \overline{H}$ \\ \hline
      \end{tabular}
    \end{center}
    \caption{The 2-to-2 processes induced by $C^{\nu \nu}$.}
    \label{tab:washoutbyC5}
\end{table}

\begin{table}[t]
  \centering
  \begin{minipage}[t]{.45\textwidth}
    \begin{center}
      \begin{tabular}{|c|c|}
        \hline
        Name: $i$ & Process  \\ \hline
        (A1) & $\overline{d_p} L_{\alpha } \to \overline{Q}_r \overline{L}_{\beta } \overline{H}$  
        \\ \hline
        (A2) & $L_{\alpha } L_{\beta } H \to d_p \overline{Q}_r$
        \\ \hline
        (A3) & $L_{\alpha } Q_r H\to d_p \overline{L}_{\beta }$ 
        \\ \hline
        (A4) & $L_{\alpha } L_{\beta } Q_r\to d_p \overline{H}$ 
        \\ \hline
        (A5) & $L_{\alpha } Q_r \to d_p \overline{L}_{\beta } \overline{H}$ 
        \\ \hline
        (A6) & $L_{\alpha } L_{\beta } \to d_p \overline{Q}_r \overline{H}$ 
        \\ \hline
        (A7) & $L_{\alpha } H \to d_p \overline{Q}_r \overline{L}_{\beta }$
        \\ \hline
        (A8) & $\overline{d_p} L_{\alpha } H\to \overline{Q}_r \overline{L}_{\beta } $ 
        \\ \hline
        (A9) & $\overline{d_p} L_{\alpha } L_{\beta }\to \overline{Q}_r \overline{H}$
        \\ \hline
        (A10) & $\overline{d_p} L_{\alpha } Q_r \to \overline{L}_{\beta } \overline{H}$ 
        \\ \hline
      \end{tabular}
    \end{center}
  \end{minipage}
  \caption{The 2-to-3 and 3-to-2 processes induced by $C^{pr}_{\alpha \beta}$.}
  \label{tab:process_LLQdH}
\end{table}
In this subsection, we calculate the collision term $\mathcal{C}$ in the right-hand side of eq.~\eqref{eq:Boltzmann1} for the Weinberg and the dim.~7 LNV operators.
Those operators change the lepton number, thereby contributing to the Boltzmann equation for $B/3 - \ell_{\alpha}$.
In the calculations, we evaluate the scattering processes at the tree level.
Also, all relevant SM particles are treated as massless because we consider temperatures higher than the electroweak scale.
As we will see later, the washout effects induced by the higher-dimensional operators are more effective in higher temperature regions, and thus, such mass effects are inefficient.

The scattering rate $\gamma_i^{eq}$ in the collision term $\mathcal{C}$ is given by the processes annihilating the initial lepton $L_{\alpha}$ caused by the LNV operators. 
Table~\ref{tab:washoutbyC5} summarize the 2-to-2 processes induced by the Weinberg operator\footnote{%
  Contributions from 1-to-3 and 3-to-1 ({\it i.e.}, three-body decay and inverse decay) processes are negligible due to phase space suppression. Similarly, for dim.~7 LNV operators, 1-to-4 and 4-to-1 process are insignificant.}.
The dim.~7 LNV operators in eq.~\eqref{eq:SMEFT_dim7} cause the 2-to-3 and 3-to-2 processes shown in table~\ref{tab:process_LLQdH}. 
Because we neglect effects from the quark masses, there is no difference among the quark flavor indices $(p,r)$.
As mentioned above, the inverse and CP-conjugated processes such as $\overline{H} \overline{H} \to L_\alpha L_\beta$, $\overline{L}_\alpha \overline{L}_\beta \to HH$, and ${H} {H} \to \overline{L}_\alpha \overline{L}_\beta$ for $(\nu 1)$ are also included in the collision term. 

By the calculations given in appendix~\ref{sec:thermal}, the sum of the thermally averaged squared amplitudes for those processes is given by 
\begin{align}
  \gamma_{\mathrm{dim.5}}^{eq} &= \frac{3 T^6}{\pi^5} |C^{\nu \nu}_{\alpha \beta}|^2, \notag \\
  \gamma^{eq}_{\mathrm{dim.7}} &= \frac{81}{2} \frac{T^{10}}{\pi^7} \sum_{pr} \Big( |C_{\alpha  \beta }^{pr}|^2 + |C_{\beta  \alpha }^{pr}|^2 -\frac{1}{2} \mathrm{Re}[C_{\alpha  \beta }^{pr} C_{\beta  \alpha }^{pr *}] \Big).
\end{align}
We note that the factor $1/(1+\delta_{\alpha \beta})$ from the phase space integral for the indistinguishable leptons in $\Delta L_{\alpha} = 2$ processes ({\it e.g.}, in $(\nu 1)$) will be canceled by the factor $\Delta (B/3 - \ell_\alpha) = -2$ in the collision term.

Consequently, we obtain the collision term as
\begin{align}
  \mathcal{C} &= \sum_{\beta} \frac{2}{T} \bigg[ \Big( \mu_{L_{\alpha }} + \mu_{L_{\beta }} + 2 \mu_H \Big) \gamma^{eq}_{\mathrm{dim.5}} +  \Big( -\mu_d + \mu_{L_{\alpha }} + \mu_{L_{\beta }} + \mu_{Q} + \mu_H \Big) \gamma^{eq}_{\mathrm{dim.7}} \bigg] \notag \\
  &= - \sum_{\beta } \Big( \eta_{\Delta (B/3 - \ell_{\alpha }) } + \eta_{\Delta( B/3 - \ell_{\beta } )} + \frac{40}{237} \eta_{\Delta(B-\ell)} \Big) \notag \\
  &\quad\qquad \times \bigg[ \frac{8 T^6}{3 \pi^5} |C^{\nu\nu}_{\alpha  \beta }|^2 + \frac{36 T^{10}}{\pi^7} \sum_{pr} \Big( |C_{\alpha  \beta }^{pr}|^2 + |C_{\beta  \alpha }^{pr}|^2 -\frac{1}{2} \mathrm{Re}[C_{\alpha  \beta }^{pr} C_{\beta  \alpha }^{pr *} ] \Big)  \bigg].
\end{align}
In the final expression, we have used the relations given by 
\begin{align}
    \mu_{L_{\alpha }} + \mu_{L_{\beta }} + 2 \mu_H &= -\mu_d + \mu_{L_{\alpha }} + \mu_{L_{\beta }} + \mu_{Q} + \mu_H \notag \\
    &= -\frac{1}{3} \Big( \mu_{B/3 - \ell_{\alpha }} + \mu_{B/3 - \ell_{\beta } } + \frac{40}{237} \mu_{B-\ell} \Big) \notag \\
  &= -\frac{4T}{9} \Big( \eta_{\Delta (B/3 - \ell_{\alpha }) } + \eta_{\Delta( B/3 - \ell_{\beta } )} + \frac{40}{237} \eta_{\Delta(B-\ell)} \Big),
\end{align}
where $\mu_{B - \ell} \equiv \sum_\alpha \mu_{B/3 - \ell_\alpha}$ and $\eta_{\Delta(B - \ell)} \equiv \sum_\alpha \eta_{\Delta(B/3 - \ell_\alpha)}$.

As one can see in the collision term, the contribution from the Weinberg operator has the mass dimension $z^{-2} T^4$, 
and that from the dim.~7 LNV operators is $z^{-4} T^4$.
Therefore, the washout by the higher-dimensional operators becomes more effective in higher temperature regions.
When we set a boundary condition at a temperature $T = T_0$, {\it i.e.}, $z = z_0$, with an initial asymmetry $\bm{\eta}_{\Delta(B/3 - \ell_\alpha)}^0 = \big(\eta_{\Delta(B/3 - \ell_1)}^0,\eta_{\Delta(B/3 - \ell_2)}^0,\eta_{\Delta(B/3 - \ell_3)}^0 \big)$, it can immediately be smeared by the higher-order operators as the temperature evolves.
Since the baryon-to-photon ratio is written by
\begin{align}
  \eta_B \equiv \eta_{\Delta B} = \frac{28}{79} \eta_{\Delta(B-\ell)},
\end{align}
the initial baryon asymmetry is also washed out by the LNV operators.
In our effective-field-theory approach, direct effects of the UV theory above the SMEFT cutoff scale are not included.
Namely, the evolution of $\eta_B$ at temperatures above the cutoff scale cannot be described.
Alternatively, we set an initial value of $\eta_B$ at $z_0=1$ and study its evolution below $\Lambda$ in the following analysis.

\begin{figure}[t]
  \centering
  \includegraphics[width=0.52\linewidth]{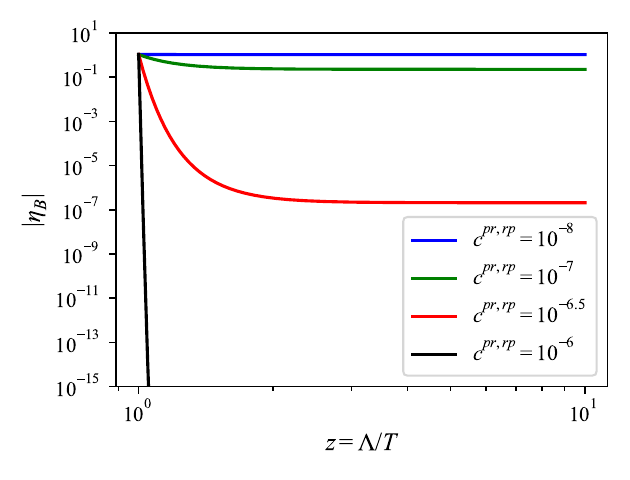}
  \includegraphics[width=0.47\linewidth]{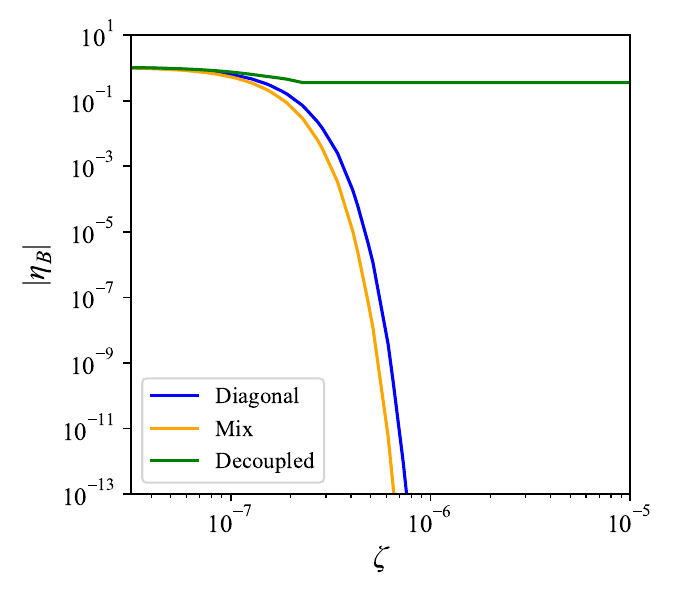}
  \caption{Left: The temperature dependence of $\eta_B$.
  The blue, green, red, and black curves correspond to $c^{pr,rp}_{\alpha \beta} = 10^{-8}$, $10^{-7}$, $10^{-6,5}$, and $10^{-6}$, respectively.
  Right: The LNV coupling dependence of $\eta_B$.
  The lepton-flavor structures are diagonal (blue), mixing (orange), and first-generation decoupled (green) cases, respectively.}
  \label{fig:leptonwashout1}
\end{figure}

Some examples are given in figure~\ref{fig:leptonwashout1}.
In these plots, we take $C^{\nu \nu}_{\alpha  \beta } = 0$ to focus on the effects coming from the dim.~7 LNV operators.
We will discuss the case with non-zero Weinberg operator in section~\ref{sec:results}.

In the left plot, the temperature dependence of $\eta_B$ is shown, where $c^{pr}_{\alpha \beta} = c^{rp}_{\alpha \beta} = 10^{-8}$ (blue), $10^{-7}$ (green), $10^{-6.5}$ (red), and $10^{-6}$ (black) are taken.
We have taken the cutoff scale as $\Lambda = 1~\mathrm{TeV}$.
We assume that the initial $B/3 - \ell_\alpha$ asymmetry is $\bm{\eta}_{\Delta(B/3- \ell_\alpha)}^0 = (1,1,1)$.
We fix the end point as $z_1 = \Lambda/T_{\mathrm{EW}}$, where $T_{\mathrm{EW}} = 10^2~\mathrm{GeV}$.
As we have mentioned above, the washout effects are more effective in higher temperature regions, and the evolution is almost constant in the low temperature region $z \simeq 10$.
The final $\eta_B$ is drastically changed between $c^{pr,rp}_{\alpha \beta} = 10^{-7}$ and $10^{-6}$.
Therefore, the final baryon asymmetry is significantly suppressed unless the initial asymmetry is extremely large for $c^{pr,rp}_{\alpha \beta} \gtrsim 10^{-6}$.

In the right panel of figure~\ref{fig:leptonwashout1}, we show the dependence of the final baryon asymmetry on the size and flavor structure of the dim.~7 couplings. 
We consider three examples with different lepton-flavor structures,
\begin{align}
  \text{diagonal:}~&c^{pr}_{\alpha \beta} = 
  \begin{pmatrix}
    \zeta & 0 & 0 \\ 0 & \zeta & 0 \\ 0 & 0 & \zeta
  \end{pmatrix}_{\alpha \beta}, ~~~~~~
  \text{mix:}~c^{pr}_{\alpha \beta} =
  \begin{pmatrix}
    0 & \zeta & 0 \\ 0 & 0 & \zeta \\ \zeta & 0 & 0
  \end{pmatrix}_{\alpha \beta}, \notag \\
  \text{decouple:}~&c^{pr}_{\alpha \beta} = 
  \begin{pmatrix}
    0 & 0 & 0 \\ 0 & 0 & \zeta \\ 0 & 0 & \zeta
  \end{pmatrix}_{\alpha \beta}.
\end{align}
Also, $c_{\alpha  \beta }^{rp} = c_{\alpha  \beta }^{pr}$ is taken.
We set the same initial asymmetry and the same cutoff scale as the left plot.
It is found from the plot that there is almost no difference between the first two cases, because the asymmetries for all lepton flavors are washed out due to the dim.~7 LNV interactions.
On the other hand, in the third case, $\eta_B$ is not smeared because the first lepton generation is decoupled from the LNV operators, and thus, the initial $B/3 - \ell_1$ asymmetry is preserved.

\section{Neutrino mass \label{sec:neutrinomass}}
\begin{figure}[t]
    \centering
    \begin{minipage}[b]{0.2\linewidth}
    \centering
    \setlength{\feynhandlinesize}{0.7pt}
    \setlength{\feynhandarrowsize}{3pt}
    \begin{tikzpicture}
    \begin{feynhand}
    \vertex (a) at (-1.2,0){\footnotesize{$\nu_{\alpha}$}}; \vertex (b) at (1.2,0){\footnotesize{$\overline{\nu_{\beta}^C}$} };
    \vertex (c) at (0,0); \vertex (d) at (0,0.8);
    \vertex (e) at (-0.4,0.4); \vertex (f) at (0.4,0.4);
    \node at (-0.65,0.2) {\scriptsize{$q_{2,r}$}};
    \node at (0.55,0.2) {\scriptsize{$d_p$}};
    \node at (0,-0.2) {\scriptsize{$C^{pr}$}};
    \node at (0,1) {\scriptsize{$u,c,t$}};
    \propag[plain] (a) to (b);
    \propag[plain] (d) to [quarter left, looseness=1] (f);
    \propag[fer] (f) to [quarter left, looseness=1] (c);
    \propag[fer] (c) to [quarter left, looseness=1] (e);
    \propag[plain] (e) to [quarter left, looseness=1] (d);
    \propag[bos] (e) to (f);
    \end{feynhand}
    \end{tikzpicture}
    \subcaption*{(A1)}
    \end{minipage}
    \begin{minipage}[b]{0.2\linewidth}
    \centering
    \setlength{\feynhandlinesize}{0.7pt}
    \setlength{\feynhandarrowsize}{3pt}
    \begin{tikzpicture}
    \begin{feynhand}
    \vertex (a) at (-0.8,0); \vertex (b) at (0.8,0);
    \vertex (c) at (0,0); \vertex (d) at (0,0.8);
    \vertex (e) at (-0.4,0.4); \vertex (f) at (0.4,0.4);
    \node at (-0.65,0.2) {\scriptsize{$q_{2,r}$}};
    \node at (0.55,0.2) {\scriptsize{$d_p$}};
    \node at (0,-0.2) {\scriptsize{$C^{pr}$}};
    \propag[plain] (a) to (b); 
    \propag[plain] (d) to [quarter left, looseness=1] (f);
    \propag[fer] (f) to [quarter left, looseness=1] (c);
    \propag[fer] (c) to [quarter left, looseness=1] (e);
    \propag[plain] (e) to [quarter left, looseness=1] (d);
    \propag[sca] (e) to (f);
    \end{feynhand}
    \end{tikzpicture}
    \subcaption*{(A2)}
    \end{minipage}
    \begin{minipage}[b]{0.2\linewidth}
    \centering
    \setlength{\feynhandlinesize}{0.7pt}
    \setlength{\feynhandarrowsize}{3pt}
    \begin{tikzpicture}
    \begin{feynhand}
    \vertex (a) at (-1.2,0){\footnotesize{$\nu_{\alpha}$}}; \vertex (b) at (1.2,0){\footnotesize{$\overline{\nu_{\beta}^C}$} };
    \vertex (c) at (0.3,0); \vertex (d) at (0.3,0.8);
    \vertex (e) at (-0.1,0.4); \vertex (f) at (0.7,0.4);
    \vertex (g) at (-0.4,0);
    \node at (0.9,0.6) {\scriptsize{$d_p$}};
    \node at (0.35,-0.2) {\scriptsize{$C^{pr}$}};
    \propag[plain] (a) to (b);
    \propag[plain] (d) to [quarter left, looseness=1] (f);
    \propag[fer] (f) to [quarter left, looseness=1] (c);
    \propag[fer] (c) to [quarter left, looseness=1] (e);
    \propag[plain] (e) to [quarter left, looseness=1] (d);
    \propag[bos] (e) to (g);
    \end{feynhand}
    \end{tikzpicture}
    \subcaption*{(B1)}
    \end{minipage}
    \begin{minipage}[b]{0.2\linewidth}
    \centering
    \setlength{\feynhandlinesize}{0.7pt}
    \setlength{\feynhandarrowsize}{3pt}
    \begin{tikzpicture}
    \begin{feynhand}
    \vertex (a) at (-0.7,0); \vertex (b) at (0.9,0);
    \vertex (c) at (0.3,0); \vertex (d) at (0.3,0.8);
    \vertex (e) at (-0.1,0.4); \vertex (f) at (0.7,0.4);
    \vertex (g) at (-0.4,0);
    \node at (0.9,0.6) {\scriptsize{$d_p$}};
    \node at (0.35,-0.2) {\scriptsize{$C^{pr}$}};
    \propag[plain] (a) to (b);
    \propag[plain] (d) to [quarter left, looseness=1] (f);
    \propag[fer] (f) to [quarter left, looseness=1] (c);
    \propag[fer] (c) to [quarter left, looseness=1] (e);
    \propag[plain] (e) to [quarter left, looseness=1] (d);
    \propag[sca] (e) to (g);
    \end{feynhand}
    \end{tikzpicture}
    \subcaption*{(B2)}
    \end{minipage}
    \caption{The Majorana neutrino mass induced by the dim.~7 LNV operators at the two-loop level.
    The quark-flavor indices are assumed to be $p \neq r$.
    The wavy and dashed lines represent the $W$ and charged NG bosons, respectively.}
    \label{fig:neutrinomass}
\end{figure}
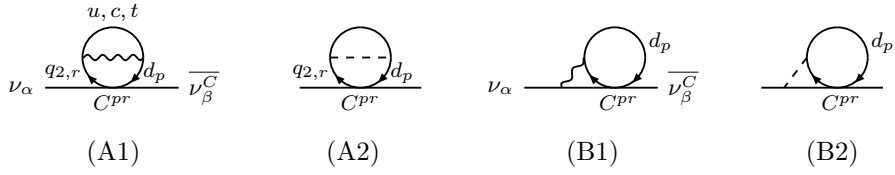

In this section, we discuss the active neutrino masses generated by the LNV operators.
The Weinberg operator generates Majorana neutrino masses at the tree level.
Moreover, the dim.~7 LNV operators affect the masses at loop levels. 
In particular, the LNV operators contributing to $b \to s \nu \nu$ and $s \to d \nu\nu$ involve quark-flavor violations and generate them at the two-loop level with electroweak interactions as shown in figure~\ref{fig:neutrinomass}.
Let us denote the active neutrino mass matrix as
\begin{align}
    m^{\nu}_{\alpha \beta} = - v^2 C_{\alpha  \beta }^{\nu \nu} - v^2 C_{\alpha  \beta }^{7},
    \label{eq:matchingforneutrinomass}
\end{align}
where the first term represents the contribution from the Weinberg operator, while those from the dim.~7 LNV operators are denoted by $C^{7}_{\alpha \beta}$, which is symmetric under $\alpha \leftrightarrow \beta$.
The neutrino mass matrix is then diagonalized by the Pontecorvo--Maki--Nakagawa--Sakata (PMNS) matrix $U$~\cite{Pontecorvo:1957qd,Maki:1962mu} as $m^\nu_{\mathrm{diag}} = U^\intercal m^{\nu}_{\alpha \beta} U$.

We represent $C^{7}_{\alpha \beta}$ in terms of the SMEFT coefficient $C^{pr}_{\alpha \beta}$ by performing the two-loop calculation explicitly.
The loop diagrams in figure~\ref{fig:neutrinomass} are classified into the two categories.
In the type-A diagrams, the quark-flavor mixing required to close the loop is caused by the one-loop sub-diagrams, in which the $W$ and charged Nambu-Goldstone (NG) bosons are involved.
On the other hand, in the B-type diagrams, the charged leptons propagate in the internal fermion lines.
Hence, the contributions to $C^{7}_{\alpha \beta}$ are denoted by
\begin{align}
    v^2 C^{7}_{\alpha \beta} = v^2 \sum_{i = 1}^2 \Big( C^{7,Ai}_{\alpha \beta} + C^{7,Bi}_{\alpha \beta} \Big),
\end{align}
where $Ai$ and $Bi$ stand for the contribution from each diagram.
The dim.~7 contributions $C^7_{\alpha \beta}$ are proportional to the quark masses because the internal quark loops involve the chirality flip via the quark masses.
We calculate these diagrams in the 't~Hooft--Feynman gauge and employ the $\overline{\mathrm{MS}}$ scheme to renormalize the UV divergences\footnote{
In the next section, we will employ the on-shell scheme to include one-loop corrections to the external fermions in the $0 \nu \beta \beta$ decay.
Here we use the $\overline{\mathrm{MS}}$ scheme for simplicity because the scheme difference arises from the higher loop order, which is expected to be small.}.
Namely, only the UV divergent parts in the two-loop functions are subtracted by the counter terms of the corresponding dim.~7 LNV operators.

For (A1), (A2), (B1), and (B2), by performing the two-loop calculations, the results are obtained as
\begin{align}
  v^2 C^{7,A1}_{\alpha \beta} &= \frac{g^2v}{\sqrt{2}} \sum_{pr} \bigg[ C_{\alpha  \beta }^{pr} m_{d_p} \big[ V^\dagger D I_1(m_u^2,m_W^2,m_{d_r}^2,m_{d_p}^2) V \big]_{rp} + (\alpha \leftrightarrow \beta) \bigg],  \\
  v^2 C^{7,A2}_{\alpha \beta} &= \frac{v}{\sqrt{2}} \sum_{pr} \bigg[ C_{\alpha  \beta }^{pr} \bigg( \big[\mathcal{Y}^{u \dagger} m_{d_p} DI_1(m_u^2,m_W^2,m_{d_r}^2,m_{d_p}^2) \mathcal{Y}^u \big]_{rp} \notag \\
  &\qquad \qquad + \big[\mathcal{Y}^{d \dagger} m_{d_p} m_{d_r} m_u I_2(m_u^2,m_W^2,m_{d_r}^2,m_{d_p}^2) \mathcal{Y}^u \big]_{rp} \notag \\
  &\qquad \qquad + \big[\mathcal{Y}^{d \dagger} m_{d_r} DI_1(m_u^2,m_W^2,m_{d_r}^2,m_{d_p}^2) \mathcal{Y}^d \big]_{rp} \notag \\
  &\qquad \qquad + \big[\mathcal{Y}^{u \dagger} m_u D I_3(m_u^2,m_W^2,m_{d_r}^2,m_{d_p}^2) \mathcal{Y}^d \big]_{rp}  \bigg) + (\alpha \leftrightarrow \beta) \bigg], \\
  v^2 C^{7,B1}_{\alpha \beta} &= \frac{g^2v}{\sqrt{2}} \sum_{pr} \bigg[ C_{\alpha  \beta }^{pr} m_{d_p} \Big[ V^\dagger DI_1(m_u^2,m_{d_p}^2,m_W^2,m_{e_{\alpha }}^2) V \Big]_{rp} + (\alpha  \leftrightarrow \beta ) \bigg], \\
  v^2 C^{7,B2}_{\alpha \beta} &= \sum_{pr} \bigg[ C^{pr}_{\alpha  \beta } m_{e_{\alpha }}^2 \bigg( m_{d_p} \Big[V^\dagger m_u I_2(m_u^2,m_{d_p}^2,m_W^2,m_{e_{\alpha }}^2) \mathcal{Y}^u \Big]_{rp} \notag \\
  &\qquad \qquad + \Big[ V^\dagger DI_1(m_u^2,m_{d_p}^2,m_W^2,m_{e_{\alpha }}^2) \mathcal{Y}^d \Big]_{rp} \notag \\
  &\qquad \qquad - \Big[ V^\dagger DI_4(m_u^2,m_{d_p}^2,m_W^2,m_{e_{\alpha }}^2) \mathcal{Y}^d \Big]_{rp} \bigg) + (\alpha  \leftrightarrow \beta ) \bigg].
  \label{eq:neutrinomass_formula}
\end{align}
We have denoted the masses of the up- (down-) type quarks and the charged leptons by $m_{u_p}$ ($m_{d_p}$) and $m_{e_\alpha}$, respectively.
Also, we have defined the matrices $\mathcal{Y}$ as 
\begin{align}
  \mathcal{Y}^u = Y^u_{\mathrm{diag}} V, ~~ \mathcal{Y}^d = -V Y^d_{\mathrm{diag}},
\end{align}
where the diagonal Yukawa matrices are related to the fermion masses, {\it i.e.}, $Y^{d}_{\mathrm{diag}} = (\sqrt{2}/v)\mathrm{diag}(m_{d_1},m_{d_2},m_{d_3})=(\sqrt{2}/v) \mathrm{diag}(m_{d},m_{s},m_{b})$, etc.
The flavor indices for the up-type quarks are abbreviated in the bracket, in which the summation is implicitly taken. 

The loop functions $I_i (x,y,z,w)$ $(i=1,...,4)$ are written in terms of the master integrals~\cite{Ford:1992pn,Martin:2001vx}, whose definitions are given in appendix~\ref{sec:twoloop}.
Since the divergences are regularized by the dimensional regularization, $D = 4 - 2 \varepsilon$ appears in the above formulae.
After the renormalization by the $\overline{\mathrm{MS}}$ scheme, these functions become finite, and they can be expressed in terms of functions of $\overline{\ln}x \equiv \ln x/Q^2$, where $Q$ is the renormalization scale.
We set the renormalization scale as $Q = \Lambda$ in the two-loop functions\footnote{
    The renormalization scale $Q$ is assumed to be not far from the electroweak scale to avoid large logarithmic corrections.
    For details of the choice of the renormalization scale, see ref.~\cite{Fridell:2024pmw}.
}. 

In the following numerical analysis, we determine the coefficient of the Weinberg operator such that the experimental results for the neutrino masses are explained. 
Namely, we treat $C^{\nu \nu}_{\alpha \beta}$ as a function of the dim.~7 coefficients $C^{pr}_{\alpha \beta}$ from eq.~\eqref{eq:matchingforneutrinomass} as 
\begin{align}
    C_{\alpha \beta }^{\nu \nu} = - C^{7}_{\alpha \beta } \pm \frac{1}{v^2} m_{\alpha \beta}^\nu \equiv \tilde{C}_{\alpha \beta }^{\nu \nu}.
    \label{eq:weinberg_def}
\end{align}
The overall sign of $m_{\alpha \beta}^\nu$ is not determined by the experiments, and we take it such that $|\tilde{C}_{\alpha \beta}^{\nu \nu}|$ is small.
Thus, in this parametrization, the experimental results can always be explained for any value of $C^{pr}_{\alpha \beta}$.

\section{Neutrinoless double beta decay \label{sec:0nuee}}

\begin{figure}[t]
  \centering
    \begin{minipage}[b]{0.49\linewidth}
    \centering
    \begin{tikzpicture}[baseline=0cm]
    \setlength{\feynhandlinesize}{0.7pt}
    \setlength{\feynhandarrowsize}{3pt}
    \begin{feynhand}
    \vertex (a) at (-0.5,0){\footnotesize{$d$}}; \vertex (d) at (1.5,0){\footnotesize{$u$}};
    \vertex (e) at (0.7,0); 
    \vertex (f) at (0.7,-1){\footnotesize{$\nu_\beta$}};
    \vertex (g) at (1.7,-0.7){\footnotesize{$e$}};
    \node at (0.8,0.25) {\scriptsize{$C^{dr}$}};
    \propag[fer] (a) to (d);
    \propag[fer] (f) to (e);
    \propag[fer] (e) to (g);
    \end{feynhand}
    \end{tikzpicture}
    \subcaption{Diagram at the tree level.}
    \end{minipage}
    %
    \begin{minipage}[b]{0.49\linewidth}
    \centering
  \begin{tikzpicture}[baseline=0cm]
    \setlength{\feynhandlinesize}{0.7pt}
    \setlength{\feynhandarrowsize}{3pt}
    \begin{feynhand}
    \vertex (a) at (-1.1,0){\footnotesize{$d$}}; \vertex (b) at (-0.6,0); \vertex (c) at (0.1,0); \vertex (d) at (1.5,0){\footnotesize{$u$}};
    \vertex (e) at (0.7,0); 
    \vertex (f) at (0.7,-1){\footnotesize{$\nu_\beta$}};
    \vertex (g) at (1.7,-0.7){\footnotesize{$e$}};
    \node at (0.4,-0.2) {\scriptsize{$d_p$}};
    \node at (0.8,0.25) {\scriptsize{$C^{pr}$}};
    \propag[fer] (a) to (d);
    \propag[boson] (b) to [half left, looseness=1.7] (c);
    \propag[fer] (f) to (e);
    \propag[fer] (e) to (g);
    \end{feynhand}
    \end{tikzpicture}
    \begin{tikzpicture}[baseline=0cm]
    \setlength{\feynhandlinesize}{0.7pt}
    \setlength{\feynhandarrowsize}{3pt}
    \begin{feynhand}
    \vertex (a) at (-1.1,0){\footnotesize{$d$}}; \vertex (b) at (-0.6,0); \vertex (c) at (0.1,0); \vertex (d) at (1.5,0){\footnotesize{$u$}};
    \vertex (e) at (0.7,0); 
    \vertex (f) at (0.7,-1){\footnotesize{$\nu_\beta$}};
    \vertex (g) at (1.7,-0.7){\footnotesize{$e$}};
    \node at (0.4,-0.2) {\scriptsize{$d_p$}};
    \node at (0.8,0.25) {\scriptsize{$C^{pr}$}};
    \propag[fer] (a) to (d);
    \propag[chasca] (b) to [half left, looseness=1.7] (c);
    \propag[fer] (f) to (e);
    \propag[fer] (e) to (g);
    \end{feynhand}
    \end{tikzpicture}
    \subcaption{Diagrams at the one-loop level.}
    \end{minipage}
    \caption{The long-range contributions to the $0 \nu \beta \beta$ decay induced by the quark-flavor-changing dim.~7 LNV operators with $p \neq r$.}
    \label{fig:0nubetabeta}
\end{figure}
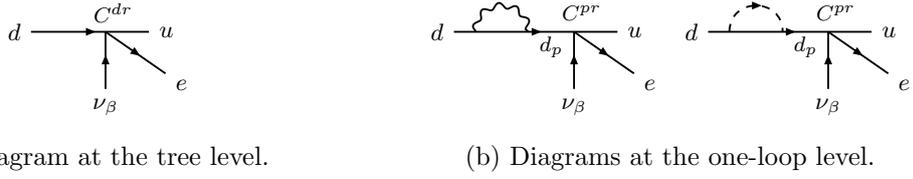

In this section, we discuss the neutrinoless double beta decay ($0 \nu \beta \beta$ decay) induced by the LNV operators.
From the latest result of KamLAND-Zen, the half-life of $^{136} \mathrm{Xe}$ was obtained as~\cite{KamLAND-Zen:2024eml}
\begin{align}
    T_{1/2}^{0 \nu, \mathrm{Xe}} > 3.8 \times  10^{26} ~\mathrm{yr}~~(90\%~\mathrm{C.L.}).
    \label{eq:Xebound}
\end{align}
In the mechanism where the neutrino mass is caused by the Majorana mass term, the effective Majorana mass $m_{ee} = |\sum_{i} m^\nu_{\mathrm{diag},i} U_{1i}^2|$ has been constrained~\cite{Vissani:1999tu}.
In the presence of the dim.~7 LNV operators, there are additional long-range contributions, as shown in figure~\ref{fig:0nubetabeta}, which are parametrized by the following effective Lagrangian~\cite{Pas:1999fc,Deppisch:2012nb},
\begin{align}
  \mathcal{L}_{\mathrm{eff}} = \frac{G_F}{\sqrt{2}} \sum_{\Gamma_a \Gamma_b} \epsilon^{\Gamma_a}_{\Gamma_b} \Big( \overline{\psi_{e}} \Gamma_a \psi_{\nu}^C \Big) \Big( \overline{\psi_{u}} \Gamma_b \psi_d \Big),
  \label{eq:eff_for_0vee}
\end{align}
where $\Gamma = \{ 1\pm \gamma^5, \gamma^\mu  (1\pm \gamma^5), \sigma^{\mu \nu} (1 + \gamma^5), \sigma^{\mu \nu} (1 - \gamma^5) \} \equiv \{ S\pm P, V \pm A, T_R, T_L \}$ are used.
The field $\psi$ represents a four-component Dirac fermion, {\it i.e.}, $\psi_u = (q_{1,1},u_1)^\intercal$ for the up quark, etc.
As we will show below, the LNV operator \eqref{eq:SMEFT_dim7} contributes to $\epsilon_{S+P}^{S+P}$ and $\epsilon_{T_R}^{T_R}$.
Following the discussions given in refs.~\cite{Deppisch:2012nb,Deppisch:2017ecm}, their experimental limits are obtained from eq.~\eqref{eq:Xebound} as
\begin{align}
  \Big| \epsilon^{S+P}_{S+P} \Big| \le 0.14 \times 10^{-8}, ~~~~\Big|\epsilon^{T_R}_{T_R} \Big| \le 0.016 \times 10^{-8}.
  \label{eq:const0vbb}
\end{align}

Let us represent $\epsilon_{S+P}^{S+P}$ and $\epsilon_{T_R}^{T_R}$ in terms of $C^{pr}_{\alpha \beta}$ ($p \neq r$).
As shown in figure~\ref{fig:0nubetabeta}, there are two types of contributions. 
In the diagram~(a), the coefficient $C^{dr}_{1\beta}$ $(r\neq d,\beta=1,2,3)$ induces the $d \to u e \nu_\beta$ transition at the tree level, because the upper component of the quark doublet depends on the CKM matrix in the mass eigenstate basis.
Here, the contribution to the $0\nu\beta\beta$ decay is summed over the lepton-flavor index $\beta$ because the neutrino flavor is unspecified in the experiments.
On the other hand, even if the external down-type quark does not take the first generation, the LNV operators can contribute to the $0\nu\beta\beta$ decay at the one-loop level, where the quark flavor is changed through self-energy corrections involving the $W$ or charged NG bosons exchange, as shown in diagram~(b).
Here, the chirality of the down-type quark is flipped, and thus, the contributions are proportional to the down-type quark masses.
We employ the on-shell scheme for the loop calculation~\cite{Aoki:1982ed,Bohm:1986rj,Denner:1990yz,Denner:1991kt,Kniehl:1996bd,Pilaftsis:2002nc,Kanemura:2004mg,Endo:2020kie} to include the flavor-changing corrections appropriately.

From eq.~\eqref{eq:SMEFT_dim7}, the SMEFT operators include the following term,
\begin{align}
  \mathcal{L}_{\mathrm{SMEFT}} &\supset \sum_{pr} \sum_{\beta} \bigg\{ -\frac{v}{\sqrt{2}} C_{\alpha \beta}^{pr *} \big[ \overline{(V^\dagger q_1)_r}  \nu_\beta^C  \big] \big[ \overline{l_\alpha} d_p^{0} \big] \bigg\} + \mathrm{h.c.},
\end{align}
where the superscript $0$ means the bare fields, {\it i.e.}, $d_p^0 = \sum_r Z^{1/2}_{R,pr} d_r = \sum_r (\delta_{pr} + \frac{1}{2} \delta Z_{R,pr})d_r$.
By taking the flavor-changing radiative corrections into account, the effective interactions contributing to the $0 \nu \beta \beta$ decay are given by 
\begin{align}
    \mathcal{L}_{\mathrm{eff}} = - \frac{v}{\sqrt{2}} \sum_{pr} \sum_{\beta} \Big( \delta_{p1} +\frac{1}{2} \delta Z_{R,p1} \Big) V_{1r} C_{1 \beta}^{pr *} \big( \overline{q_{1,1}} \nu_\beta^C \big) \big( \overline{l_1} d_1 \big) + \mathrm{h.c.}.
\end{align}
By matching them onto eq.~\eqref{eq:eff_for_0vee}, $\epsilon_{S+P}^{S+P}$ and $\epsilon_{T_R}^{T_R}$ are expressed as
\begin{align}
  \epsilon^{S+P}_{S+P} &= \frac{\sqrt{2} v^3}{8} \sum_{pr} \Big( \delta_{p1} + \frac{1}{2}\delta Z_{R,p1} \Big)  V_{1r} \sum_\beta C^{pr*}_{1 \beta} , \notag \\
  \epsilon^{T_R}_{T_R} &= \frac{\sqrt{2} v^3}{32} \sum_{pr} \Big( \delta_{p1} + \frac{1}{2}\delta Z_{R,p1} \Big) V_{1r} \sum_\beta C^{pr*}_{1 \beta} .
  \label{eq:0vbb_loop_tree} 
\end{align}

In the on-shell scheme, the renormalized quark two-point functions are required to coincide with the tree-level propagators at their pole positions, with no flavor mixing and unit residues.
As a result, the two-point functions remain flavor diagonal at their pole positions even at the loop level, and the quark-flavor-changing effects are encoded in the off-diagonal wave-function renormalization constants of the quarks.
For the diagonal part of the wave-function renormalization factors, the corrections are given by
\begin{align}
  \delta Z_{R,pp} = - \Sigma_{pp}^R (m_{d_p}^2) + A_{p} -D_{p},
\end{align}
where the right-handed side is
\begin{align}
  &A_p = \frac{1}{2m_{d_p}} \Big( \Sigma_{pp}^{DR}(m_{d_p}^2) - \Sigma_{pp}^{DL}(m_{d_p}^2) \Big), \notag \\
  &D_p = m_{d_p}^2 \Big( \Sigma_{pp}^{L \prime} (m_{d_p}^2) + \Sigma_{pp}^{R \prime} (m_{d_p}^2) \Big) + m_{d_p} \Big( \Sigma_{pp}^{DL \prime} (m_{d_p}^2) + \Sigma_{pp}^{DR \prime} (m_{d_p}^2) \Big).
\end{align}
The prime denotes the derivative with respect to the external momentum $p^2$ at $p^2 = m_{d_p}^2$.
For the off-diagonal part, we obtain
\begin{equation}
  \delta Z_{R,pr} = \frac{-2}{m_{d_r}^2 - m_{d_p}^2} \Big[ m_{d_r}^2 \Sigma_{pr}^R (m_{d_r}^2) + m_{d_p} m_{d_r} \Sigma_{pr}^L (m_{d_r}^2) + m_{d_p} \Sigma^{DR}_{pr} (m_{d_r}^2) + m_{d_r} \Sigma^{DL}_{pr} (m_{d_r}^2)  \Big],
  \label{eq:renfac_offd}
\end{equation}
where the self-energy functions are written as 
\begin{align}
  &\Sigma^L_{pr}(p^2) = - \frac{1}{16\pi^2} \bigg\{ g^2 \Big[V^\dagger B_1[p^2;m_u^2,m_W^2] V \Big]_{pr} + \Big[\mathcal{Y}^{u \dagger} B_1[p^2;m_u^2,m_W^2] \mathcal{Y}^u \Big]_{pr} \bigg\}, \notag \\
  &\Sigma^R_{pr}(p^2) = - \frac{1}{16\pi^2} \Big[\mathcal{Y}^{d \dagger} B_1[p^2;m_u^2,m_W^2] \mathcal{Y}^d \Big]_{pr}, \notag \\
  &\Sigma^{DL}_{pr}(p^2) = \frac{1}{16\pi^2} \Big[\mathcal{Y}^{d \dagger} m_u B_0[p^2;m_u^2,m_W^2] \mathcal{Y}^u \Big]_{pr}, \notag \\
  &\Sigma^{DR}_{pr}(p^2) = \frac{1}{16\pi^2} \Big[\mathcal{Y}^{u \dagger} m_u B_0[p^2;m_u^2,m_W^2] \mathcal{Y}^d \Big]_{pr}.
\end{align}
The analytic formulae of the Passarino--Veltmann functions~\cite{tHooft:1978jhc,Passarino:1978jh}, $B_0$ and $B_1$, are given in appendix~\ref{sec:twoloop}.
We note that the self-energy diagrams depend on the renormalization scale $Q$.
Since the dependence is logarithmic, the result is not sensitive to the choice of this value.
In the following analysis, we take $Q = m_W$.

As we can see in eq.~\eqref{eq:0vbb_loop_tree}, the $S+P$ contribution from the dim.~7 LNV operators is four times larger than that of $T_R$.
However, as in eq.~\eqref{eq:const0vbb}, the constraint on $\epsilon_{T_R}^{T_R}$ is about ten times stronger than that on $\epsilon_{S+P}^{S+P}$.
Therefore, we simply apply the latter stronger condition to the upper bound on the LNV operators in the next section.

\section{Results \label{sec:results}}

In this section, we show the numerical results for the dim.~7 LNV operators relevant to the $B \to K \nu \nu$ and $K \to \pi \nu \nu$ decays.
We consider the following three scenarios:
\begin{itemize}
  \item \textbf{Scenario 1}
  
  In this scenario, we take flavor-universal Wilson coefficients as
  \begin{align}
    c^{pr}_{\alpha \beta} = c^{rp}_{\alpha \beta} \equiv \zeta
    \begin{pmatrix}
      1 & 1 & 1 \\
      1 & 1 & 1 \\
      1 & 1 & 1 
    \end{pmatrix}_{\alpha \beta},
    \label{eq:scenario1}
  \end{align}
  where $\zeta$ is a constant.
  \item \textbf{Scenario 2}
  
  As we have seen in section~\ref{sec:neutrinomass}, the neutrino masses induced by the dim.~7 LNV operators are proportional to $C^{pr}_{\alpha \beta}$.
  In this scenario, we require that its lepton-flavor structure is proportional to the experimental results for the neutrinos as 
  \begin{align}
    c^{pr}_{\alpha \beta} = c^{rp}_{\alpha \beta} = \zeta m^{\nu}_{\alpha \beta}.
  \end{align}
  \item \textbf{Scenario 3}
  
  In this scenario, we assume that the dim.~7 LNV interactions for a specific lepton flavor is suppressed as 
  \begin{align}
    c^{pr}_{\alpha \beta} = c^{rp}_{\alpha \beta}  = \zeta 
    \begin{pmatrix}
      r & r & r \\
      r & 1 & 1 \\
      r & 1 & 1 
    \end{pmatrix}_{\alpha \beta}.
    \label{eq:scenario3}
  \end{align}
  For $r \ll 1$, the first generation is decoupled.
  Consequently, as we have discussed in section~\ref{sec:washout}, the initial lepton asymmetry of that flavor would not be washed out in the limit of $r \to 0$.
\end{itemize}
In the following, we take the normal ordering for the neutrino mass spectrum and assume that the lightest neutrino mass is zero.
Also, the NuFIT5.2 results with SK-ATM data are adopted~\cite{Esteban:2020cvm,NuFIT}.

\subsection{Analysis for \texorpdfstring{$B \to K \nu \nu$}{B to K nu nu}}
In this subsection, we show the numerical results for $C^{sb,bs}_{\alpha \beta}$, which are related to the $B^+ \to K^+ \nu \nu$ decay.

\begin{figure}[t]
  \centering
  \includegraphics[width=0.45\linewidth]{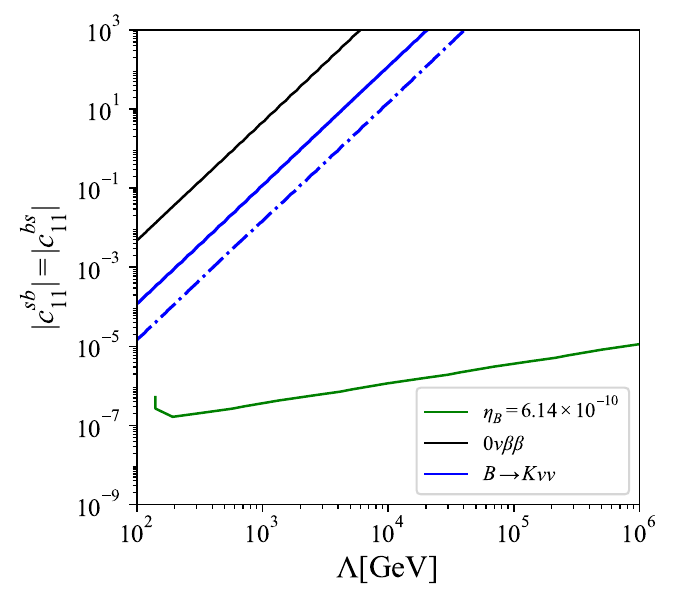}
  \includegraphics[width=0.45\linewidth]{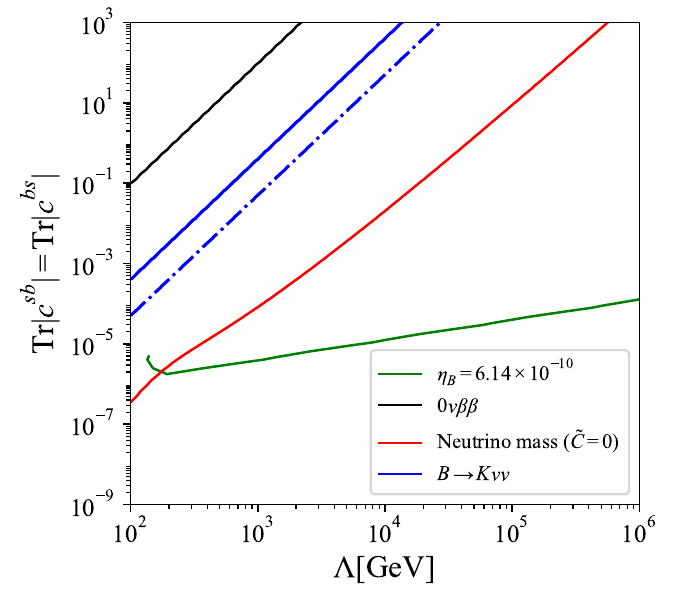}
  \caption{Left (right): The results for varying $C^{sb}_{\alpha \beta}$ and $C^{bs}_{\alpha \beta}$ in Scenario 1 (Scenario 2).
  The blue solid (dash-dotted) lines show the current bound (future projection) of the $B^+ \to K^+ \nu \nu$ decay.
  The black lines are the constraints from the $0 \nu \beta \beta$ decay.
  The primordial baryon asymmetry is washed out by the LNV operators, and the corresponding regions for $\eta_B = 6.14 \times 10^{-10}$ are denoted by the green lines.
  In the right plot, the red line represents the case when the neutrino masses are explained only by the dim.~7 LNV operators.}
  \label{fig:scenario1and2}
\end{figure}
\begin{figure}[t]
  \centering
  \includegraphics[width=0.45\linewidth]{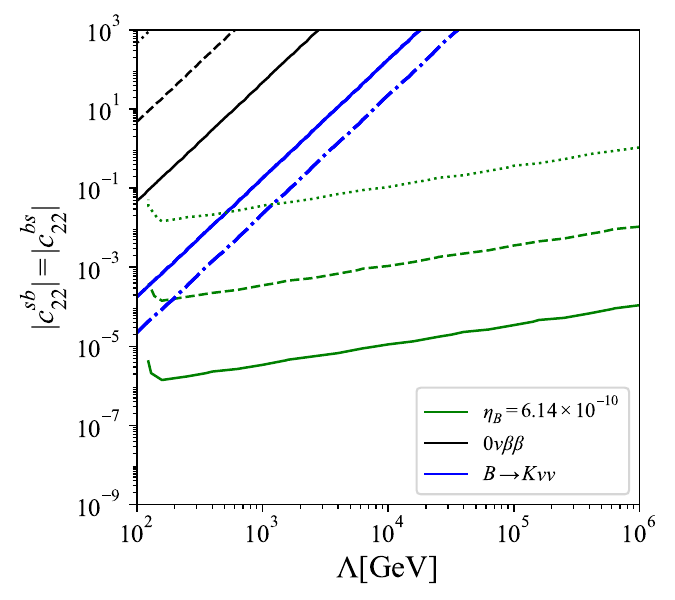}
  \includegraphics[width=0.53\linewidth]{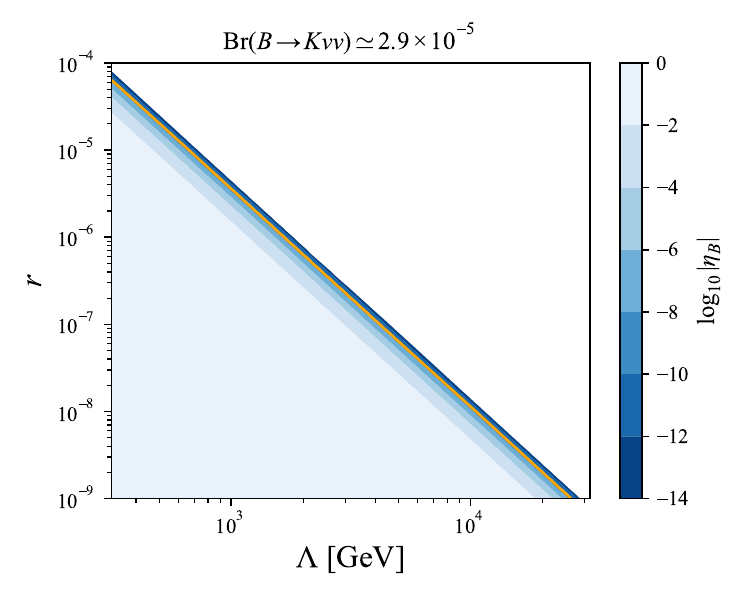}
  \caption{Left: Same as figure~\ref{fig:scenario1and2} in Scenario 3 with $r = 10^{-1}$ (solid), $r=10^{-3}$ (dashed), and $r=10^{-5}$ (dotted).
  Right: the contour plot of $\eta_B$ in the $r$-$\Lambda$ plane.
  The orange line shows $\eta_B = 6.14 \times 10^{-10}$.}
  \label{fig:scenario3}
\end{figure}

In figure~\ref{fig:scenario1and2}, the upper limits allowed by $B^+ \to K^+ \nu \nu$ from eq.~\eqref{eq:BKvvbound} are shown by the blue solid lines in the $c^{sb,bs}_{\alpha \beta}$-$\Lambda$ plane.
The blue dash-dotted lines indicate the projected 95\% C.L.~bounds expected at Belle II with $50$~$\mathrm{ab}^{-1}$, where we assume that the central value of the measurement coincides with the SM prediction.
The left (right) panel corresponds to Scenario 1 (Scenario 2).
In the right panel, the vertical axis is the trace over the lepton flavor indices.
The baryon asymmetry is evolved from $T= \Lambda$ to $T_{\mathrm{EW}}$ with the initial condition $\bm{\eta}_{\Delta(B/3-\ell_\alpha)}^0= (1,1,1)$.
The green curves correspond to $\eta_B = 6.14 \times 10^{-10}$~\cite{Planck:2018vyg}.
In the regions above them, the initial baryon asymmetry is smeared too much, and the final baryon asymmetry is too small to explain the observed value.
The black lines are the upper bounds from the $0 \nu \beta \beta$ decay.

The red curve in the right plot represents the region where $\tilde{C}_{\alpha \beta}^{\nu \nu} = 0$, indicating that the neutrino mass is explained purely by the dim.~7 LNV operators.
It is noted that the neutrino mass itself does not directly constrain the scenario.
Away from this curve, the Weinberg operator becomes non-zero and explains the neutrino mass.
Although this operator could, in principle, affect the baryon number washout, we found that such effects are sufficiently small within the parameter region shown in the plot. 
On the other hand, the contribution of the Weinberg operator to the $0 \nu \beta \beta$ decay is effectively hidden because it appears only via the neutrino mass, which is fixed by the experimental data.

As seen from Scenarios 1 and 2, in the regions where the dim.~7 LNV operators cause sizable effects on the $B^+ \to K^+ \nu \nu$ decay, the initial baryon asymmetry is washed out significantly, and the scenarios are unlikely to survive.
However, if the cutoff scale $\Lambda$ is as low as $100~\mathrm{GeV}$, the baryon number washout due to the LNV operators is inefficient because the time interval of the Boltzmann equation, {\it i.e.}, the range between the cutoff and electroweak scales, is small. 
Then, the washout above the cutoff scale in the UV theory should be taken into account, which is beyond the scope of this paper.
It is also found in the Scenario 2 that the neutrino mass is dominated by the Weinberg operator in the region allowed by the baryon asymmetry, {\it i.e.}, below the green line.

In the left panel of figure~\ref{fig:scenario3}, the result for Scenario 3 is shown.
The vertical axis is $|c_{22}^{sb}| = |c_{22}^{bs}|$, which are the dim.~7 LNV couplings for the second generation of the leptons.
We consider the case where the first generation is decoupled from the LNV interactions in the limit of $r \to 0$ (see eq.~\eqref{eq:scenario3}).
The solid, dashed, and dotted lines correspond to $r=10^{-1},~10^{-3}$, and $10^{-5}$, respectively.

It is found that the decay rate of $B^+ \to K^+ \nu \nu$ is almost unchanged by $r$ because the rate is dominated by the decays into the second- and third-generation neutrinos.
On the other hand, the $0 \nu \beta \beta$ bounds are sensitive to $r$.
This is because only the LNV operators of the first generation are constrained by the $0 \nu \beta \beta$ decay, and in the limit of $r \to 0$, there is no long-range contribution from $C^{sb}_{\alpha \beta}$ and $C^{bs}_{\alpha \beta}$.

In addition, the upper limit from the baryon asymmetry is sensitive to $r$.
As we can read from the left panel of figure~\ref{fig:scenario1and2}, $\eta_{\Delta (B/3-\ell_2)}$ and $\eta_{\Delta (B/3-\ell_3)}$ are washed out efficiently for $|c_{22}^{sb,bs}| \gtrsim 10^{-(5-7)}$ depending on $\Lambda$.
Thus, in the left panel of figure~\ref{fig:scenario3}, the baryon asymmetry is dominated by $\eta_{\Delta (B/3-\ell_1)}$, and its washout is governed by $r \times |c_{22}^{sb,bs}|$, resulting in the green lines.
For $\Lambda \gtrsim 200~\mathrm{GeV}$, there are the regions where the LNV contributions to $\mathrm{Br}(B^+ \to K^+ \nu \nu)$ becomes sizable and can be probed in the near future at Belle~II avoiding the constraint from the baryon number washout if $r$ is less than about $10^{-3}$.
For smaller $\Lambda$, the baryon number washout is weakened.

In Scenario~3, we have considered the case when the interaction with the first generation of the leptons is suppressed in the dim.~7 LNV operators.
Alternatively, if we consider the cases when $c^{bs,sb}_{\alpha \beta}$ for the second or third generation leptons are suppressed, the results for $B^+ \to K^+ \nu \nu$ and the baryon asymmetry are unchanged from figure~\ref{fig:scenario3} just by replacing the axis as $|c^{sb,bs}_{22}| \to |c^{sb,bs}_{11}|$ in the left panel.
On the other hand, the $0 \nu \beta \beta$ bound becomes the same as that of the left plot of figure~\ref{fig:scenario1and2} and is insensitive to $r$, because only $c^{sb}_{1 \beta}$ and $c^{bs}_{1 \beta}$ are constrained.
We find that, even in such cases, $r \lesssim 10^{-3}$ is required for the scenario of the sizable LNV contributions to $\mathrm{Br}(B^+ \to K^+ \nu \nu)$ to work consistently satisfying the constraints.

Finally, the right panel of figure~\ref{fig:scenario3} shows contours of $\eta_{B}$ in the $r$-$\Lambda$ plane.
The orange line corresponds to $\eta_{B} = 6.14 \times 10^{-10}$.
The value of $\zeta$ is determined such that the branching ratio becomes $\mathrm{Br}(B^+ \to K^+ \nu \nu) \simeq 2.9 \times 10^{-5}$, which corresponds to $C^{sb,bs}_{22} = 10^{-10}$~$\mathrm{GeV}^{-3}$ and to the middle region of the blue solid and dash-dotted lines in the left panel.
In this case, $\eta_{\Delta (B/3-\ell_2)}$ and $\eta_{\Delta (B/3-\ell_3)}$ are washed out sufficiently, and the baryon asymmetry is determined by $\eta_{\Delta (B/3-\ell_1)}$.
As $r$ decreases, more baryon asymmetry survives, and the initial $\eta_{\Delta (B/3-\ell_1)}$ is conserved in the limit of $r \to 0$.
Therefore, when the dim.~7 LNV operators give sizable contributions to the $B^+ \to K^+ \nu \nu$ decay, $r$ should be suppressed, {\it e.g.}, $r \lesssim 10^{-5}$ for $\Lambda = 10^3~\mathrm{GeV}$, to avoid the washout of the baryon asymmetry.

Since we introduce the Weinberg operator to explain the neutrino mass as in eq.~\eqref{eq:weinberg_def}, the first generation of the lepton number can be changed via the processes shown in table~\ref{tab:washoutbyC5}.
However, such an effect is negligibly small due to the smallness of the neutrino masses.

\subsection{Analysis for \texorpdfstring{$K \to \pi \nu \nu$}{K to pi nu nu}}

\begin{figure}[t]
  \centering
  \includegraphics[width=0.45\linewidth]{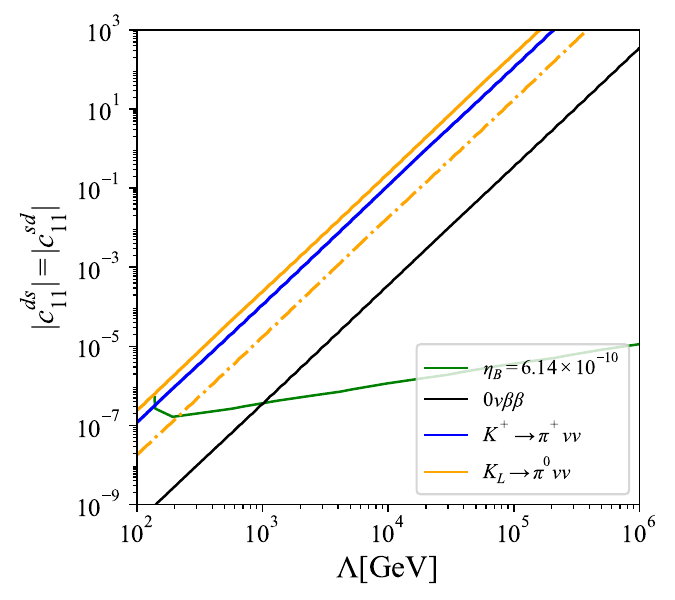}
  \includegraphics[width=0.45\linewidth]{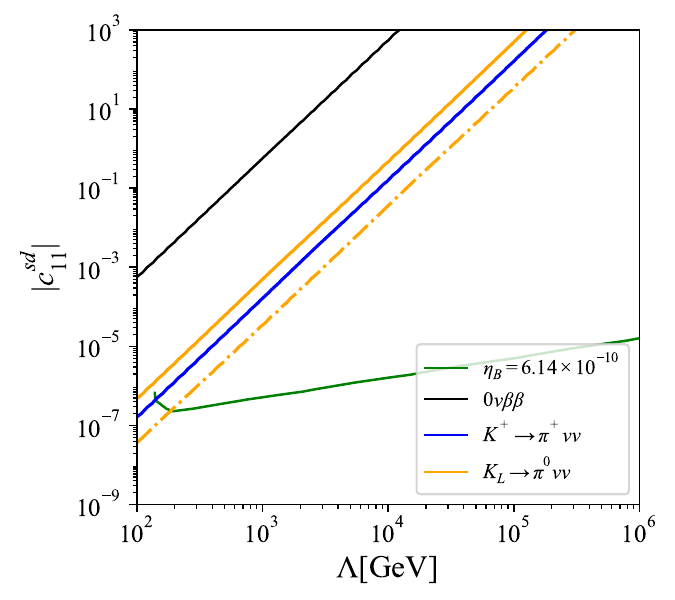}
  \caption{Left: The result for varying $C^{ds}$ and $C^{sd}$ in Scenario 1. 
  The blue and orange solid lines are the experimental upper bounds from the $K^+ \to \pi^+ \nu \nu$ and $K_L \to \pi^0 \nu \nu$ decays, respectively.
  The orange dash-dotted lines are the future projections of $K_L \to \pi^0 \nu \nu$ at KOTO~II.
  The black and green lines show the upper bound from the $0 \nu \beta \beta$ decay and the region corresponding to $\eta_B = 6.14 \times 10^{-10}$, respectively.
  Right: the case obtained by taking $C^{ds} = 0$ from the left plot.}
  \label{fig:K_scenario1}
\end{figure}
\begin{figure}[t]
  \centering
  \includegraphics[width=0.45\linewidth]{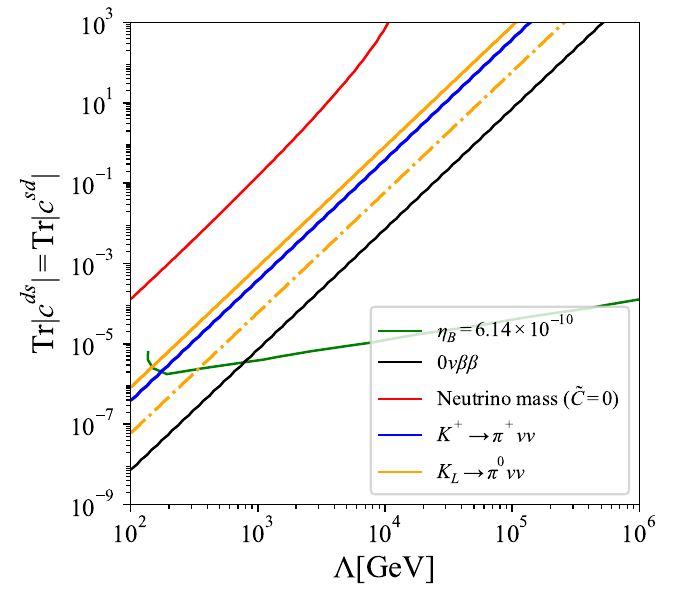}
  \includegraphics[width=0.45\linewidth]{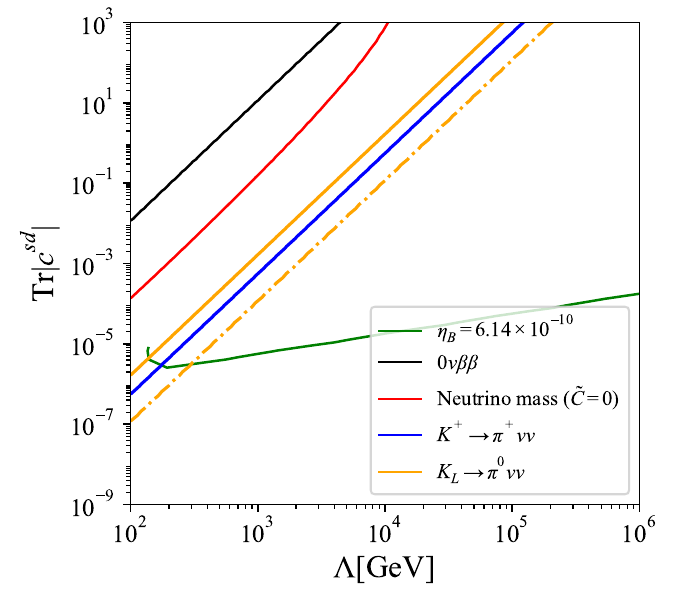}
  \caption{Same as in figure~\ref{fig:K_scenario1} in Scenario 2.
  As in the right plot in figure~\ref{fig:scenario1and2}, the red line corresponds to the case when the neutrino masses are explained by the dim.~7 LNV operators.}
  \label{fig:K_scenario2}
\end{figure}

Similarly to the previous subsection, we analyze $C^{ds,sd}_{\alpha \beta}$ for the $K \to \pi \nu \nu$ decays.

In figure~\ref{fig:K_scenario1}, the lepton-flavor structure is given by Scenario 1, though $C^{ds}_{\alpha \beta}$ is turned off in the right panel. 
The blue line corresponds to the $2 \sigma$ upper bound from $K^+ \to \pi^+ \nu \nu$ decays reported by the NA62 experiment~\cite{NA62:2024pjp}. 
The orange line shows the $90 \%~\mathrm{C.L.}$ upper bounds from the $K_L \to \pi^0 \nu \nu$ decay at KOTO~\cite{KOTO:2024zbl}.
The orange dash-dotted lines correspond to a $40\%$ deviation from the SM central value, which would be probed by KOTO II with $90\%$ significance~\cite{KOTO:2025uqg}.
The contours of $\eta_B$ shown by the green lines are not changed from the left panel of figure~\ref{fig:scenario1and2}.
In contrast, the constraint from the $0 \nu \beta \beta$ decay is sensitive to $C^{ds}_{\alpha \beta}$.
As shown by the black line in the left panel, a wider parameter space is limited than those in figure~\ref{fig:scenario1and2} because the coupling $C^{ds}_{\alpha \beta}$ causes the decay at the tree level.
On the other hand, in the right panel, the $0 \nu \beta \beta$ constraints are relaxed due to $C^{ds}_{\alpha \beta} = 0$, where the decay is induced at the one-loop level.

\begin{figure}[t]
  \centering
  \includegraphics[width=0.45\linewidth]{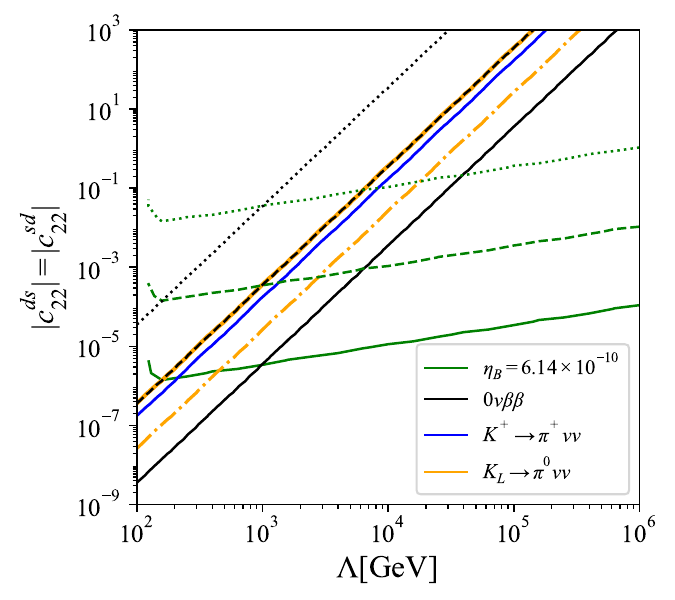}
  \includegraphics[width=0.45\linewidth]{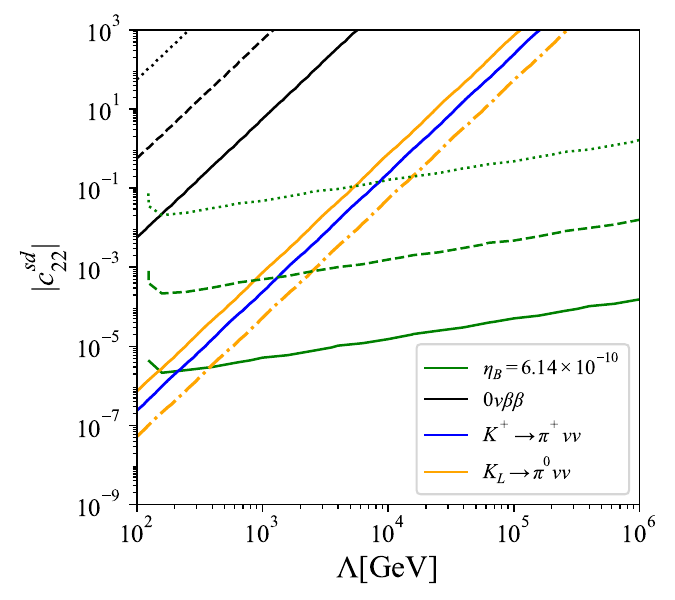}
  \caption{Same as figure~\ref{fig:K_scenario1} in Scenario 3.
  The solid, dashed, and dotted lines are for the case with $r = 10^{-1},10^{-3}$, and $10^{-5}$, respectively.}
  \label{fig:K_scenario3}
\end{figure}

Similar plots for Scenario 2 are shown in figure~\ref{fig:K_scenario2}.
Again, $C^{ds}_{\alpha \beta} = 0$ is taken in the right plot.
The red lines show the regions where the neutrino mass is explained only by the dim.~7 LNV operators.
The loop contributions to the neutrino mass depend on the CKM matrix and the quark masses.
The red lines appear in the region excluded by both $K^+ \to \pi^+ \nu \nu$ and $K_L \to \pi^0 \nu \nu$ decays.
Meanwhile, the results for Scenario~3 are shown in figure~\ref{fig:K_scenario3}.
The differences from figure~\ref{fig:scenario3} can be understood similarly to figure~\ref{fig:K_scenario1}.

From the results for Scenario~1 and 2, the setup with non-zero $C^{ds}$ is strongly constrained by the $0\nu \beta \beta$ decay.
In addition, even if $C^{ds} = 0$, the regions of large $\Lambda$ are constrained to avoid the baryon number washout.
Therefore, for the measurement of the sizable LNV contributions to $\mathrm{Br}(K^+ \to \pi^+ \nu \nu)$ and $\mathrm{Br}(K_L \to \pi^0 \nu \nu)$ at the future experiments, $C^{ds} = 0$ and $\Lambda \lesssim 200~\mathrm{GeV}$ are needed in Scenario~1 and 2.
On the other hand, as shown in figure~\ref{fig:K_scenario3} (Scenario~3), even for $C^{ds} \neq 0$, the constraint from the $0 \nu \beta \beta$ decay can be relaxed as long as the dim.~7 LNV couplings of the first lepton generation are suppressed.
In this case, the initial $O(1)$ asymmetry of the first-generation leptons can survive.

In Scenario~3, the interactions with the first generation of the leptons are assumed to be suppressed. 
If we consider the cases when the second or third generation is decoupled from the LNV operators, the results for the K meson decays and the baryon number washout are unchanged from those in figure~\ref{fig:K_scenario3}, {\it i.e.}, the upper bounds are obtained by replacing the axis as $|c^{ds,sd}_{22}| \to |c^{ds,sd}_{11}|$ in the left plot of figure~\ref{fig:K_scenario3}.
However, the $0 \nu \beta \beta$ limit is strong enough to constrain the scenario, as seen in the left plot of figure~\ref{fig:K_scenario1}.
For the case with $C^{ds}_{\alpha \beta} = 0$, the $0 \nu \beta \beta$ bound in this plane is much relaxed as same as in the right panel of figure~\ref{fig:K_scenario1}.
As a result, if the LNV couplings for the second or third generation are sufficiently suppressed, the constraint from the baryon number washout in this plane is weakened, as in the right plot of figure~\ref{fig:K_scenario3}.

\section{Discussion and conclusions \label{sec:discussions}}

In this paper, we have studied the LNV interactions that contribute to the semileptonic meson decays $B \to K \nu \nu$ and $K \to \pi \nu \nu$.
They are given by the $\overline{d}LQLH$-type operator at dim.~7 within the SMEFT.
Since neutrinos are not identified directly in the detectors at the flavor experiments, measurements of $\mathrm{Br}(B \to K + \cancel{E})$ and $\mathrm{Br}(K \to \pi + \cancel{E})$ are affected if the lepton number of any generation is significantly violated by the interactions.

By comparing those contributions with the constraints from the washout of the primordial baryon asymmetry of the Universe, as well as those from the $0\nu\beta\beta$ decay, we have clarified the conditions under which excesses in the meson decay rates over the SM predictions can be observed. 
For $\mathrm{Br}(B^+ \to K^+ + \cancel{E})$, we have shown that an excess could be successfully detected at Belle~II with $50$~$\mathrm{ab}^{-1}$ if at least one of the lepton flavors is almost decoupled from the LNV interactions. 
In this case, the bayon number washout constraint provides the relevant condition, while the limit from the $0\nu\beta\beta$ decay is satisfied in the parameter regions of interest. 
On the other hand, an excess could be observed for $\mathrm{Br}(K_L \to \pi^0 + \cancel{E})$ at KOTO II in the future if both electron and electron-like neutrinos are decoupled from the LNV operator to satisfy both constraints.
Alternatively, the scenario may remain viable if the LNV coupling $C^{ds}_{\alpha \beta}$ is suppressed, and if the cutoff scale $\Lambda$ is close to the electroweak scale.

In this paper, we have also analyzed the LNV contributions to the neutrino masses.
Although the masses are generated at the two-loop level from the quark-flavor changing LNV interactions, the experimental results of the neutrino masses can be explained by additionally introducing the Weinberg operator, which violates the lepton number at dim.~5.
We have studied the contributions of the Weinberg operator to the baryon number washout and to the $0\nu\beta\beta$ decay, and have found that those effects are negligibly small. 

We have assumed that an initial $B/3 - \ell_\alpha$ asymmetry is generated by some baryogenesis mechanism, such as leptogenesis~\cite{Fukugita:1986hr,Barbieri:1999ma,Davidson:2002qv,Abada:2006fw,Abada:2006ea}, and studied their washout by the LNV interactions. 
One may consider alternative mechanisms to explain the baryon asymmetry of the Universe that can be safe from the washout. 
A representative scenario is electroweak baryogenesis~\cite{Kuzmin:1985mm}, in which the baryon asymmetry is generated around the expanding bubble walls during the electroweak phase transition, where a non-equilibrium situation for the sphaleron processes is realized. 
In this case, since the baryon asymmetry is created just before the sphaleron processes freeze out, the washout by the dim.~7 LNV operators would not work efficiently. 
Similarly, other possibilities include scenarios for baryogenesis operating after sphaleron decoupling (for review, see ref.~\cite{Elor:2022hpa}).

The SMEFT operators, including the $\overline{d}LQLH$-type operators, are obtained by integrating out heavy degrees of freedom in the UV theory. 
Hence, heavy particles in the UV theory should violate lepton numbers and, in addition, have interactions that connect the lepton and quark sectors.
Such particles may be provided {\it e.g.}, in leptoquark models~\cite{Fridell:2024pmw}.
They could wash out the primordial baryon asymmetry of the Universe at the temperature above the cutoff scale of the SMEFT, and might generate other types of LNV interactions within the SMEFT, inducing additional contributions to the $0\nu\beta\beta$ decay.
Since such contributions are likely to spoil the scenarios, the results in this paper are considered to be conservative. 
However, the evaluations of those effects are beyond the scope of this paper, and further studies are required to clarify the conditions under which the LNV meson decays would be observed. 
Nonetheless, any observation of such decays could open a new direction to reveal the mystery of the lepton number violations.

\section*{Acknowledgment}
This work is supported by JSPS KAKENHI Grant Numbers 22K21347 [M.E. and Y.M.], 21K13923 [K.Y.], National Science and Technology Council Grant Number~112--2112--M--110--010 [S.I.], and by the FORTE project
CZ.02.01.01/00/22\_008/0004632 co-funded by the EU and the Ministry
of Education, Youth and Sports of the Czech Republic [K.F.].
We thank Martin Aria Mojahed for helpful discussions.

\appendix

\section{Fierz transformation \label{sec:fierz}}

By substituting $\langle H^n \rangle = v \delta_{2n} / \sqrt{2}$ into eq.~\eqref{eq:SMEFT_dim7}, we obtain
\begin{align}
  \mathcal{L} &= \frac{v}{\sqrt{2}} \sum_{p r} \sum_{\alpha \beta}  C_{\alpha \beta}^{pr} \bigg\{ (\overline{d_p} \nu_\alpha) (\overline{q_{2,r}^C} \nu_\beta) - (\overline{d_p} l_\alpha) \big(\overline{V^\dagger q_{1,r}^C} \nu_\beta \big) \Big\}  + \mathrm{h.c.}.
\end{align}
The Fierz transformation is given by 
\begin{align}
  (\psi_1 \psi_2) (\psi_3 \psi_4) &= \psi_1^a \psi_{2b} \psi_3^c \psi_{4 d} \delta^b_a \delta_c^d \notag \\
   &= \frac{1}{4} \Big[ \delta_a^d \delta_c^b + (\gamma_5)_a^d (\gamma_5)_c^b + (\gamma^\mu)_a^d (\gamma_\mu)_c^b
   \notag \\
   &\quad -(\gamma^\mu \gamma_5)_a^d (\gamma_\mu \gamma_5 )_c^b + \frac{1}{2} (\sigma^{\mu \nu})_a^d (\sigma_{\mu \nu})_c^b \Big] \psi_1^a \psi^{c \intercal}_{4} \psi_{2b}\psi_{3d}^\intercal,
\end{align}
where the summation for the spinor indices $a,b,c,d$ are implicitly taken.
Hence, we get the result,
\begin{align}
  \mathcal{L} = &-  \frac{v}{4 \sqrt{2}} \sum_{p r} \sum_{\alpha \beta} \Big\{ (C_{\alpha \beta}^{pr} + C_{\beta \alpha}^{pr})(\overline{d_p} q_{2,r}) (\overline{\nu_\alpha^C} \nu_\beta) - \frac{1}{4} (C_{\alpha \beta}^{pr} - C_{\beta \alpha}^{pr}) (\overline{d_p} \sigma^{\mu \nu} q_{2,r}) (\overline{\nu_\alpha^C} \sigma_{\mu \nu} \nu_\beta) \Big\} \notag \\
  &+ \frac{v}{2 \sqrt{2}} \sum_{p r} \sum_{\alpha \beta} \Big\{ C_{\alpha \beta}^{pr} \big( \overline{d_p} (V^\dagger q_{1})_r \big) (\overline{l_\alpha^C} \nu_\beta) - \frac{1}{4} C_{\alpha \beta}^{pr} \big( \overline{d_p} \sigma^{\mu \nu}(V^\dagger q_{1})_r \big) (\overline{l_\alpha^C} \sigma_{\mu \nu} \nu_\beta) \Big\} + \mathrm{h.c.},
\end{align}
where $\overline{\nu_\alpha^C} \nu_\beta  = \overline{\nu_\beta^C} \nu_\alpha$ and $\overline{\nu_\alpha^C} \sigma_{\mu \nu} \nu_\beta  = - \overline{\nu_\beta^C} \sigma_{\mu \nu} \nu_\alpha$ are used in the first line.
By comparing the first line and eq.~\eqref{eq:LEFT}, we obtain eq.~\eqref{eq:matching_SMEFT_LEFT}.

\section{Form factors \label{sec:formfactor}}

Following refs.~\cite{Gubernari:2018wyi,Felkl:2021uxi}, we define the form factors $f^B_0$,$f^B_+$ and $f^B_T$ used in decay processes with $B\to K$ transition as
\begin{align}
  &\bra{K (k)} \overline{s} \gamma^\mu b \ket{B (p)} = \bigg[ (p+k)^\mu - \frac{m_{B}^2 - m_K^2}{q^2} q^\mu \bigg] f^B_+(q^2) + \frac{m_{B}^2 - m_K^2}{q^2} q^\mu f^B_0(q^2), \notag \\
  &\bra{K (k)} \overline{s} \sigma^{\mu \nu} q_\nu b \ket{B (p)} = \frac{if^B_T(q^2)}{m_B + m_K} \big(q^2 (p+k)^\mu - (m_B^2 - m_K^2) q^\mu \big)
  \label{eq:Bmesonform}
\end{align}
where $p^\mu$ ($k^\mu$) is the four momentum of $B$ $(K)$ and $q^2 = (p-k)^2$.
Those for processes with the $K \to \pi$ transition are defined similarly.

The $J$ factors for the $B^+ \to K^+ \nu \nu$ decay are given by~\cite{Buras:2024ewl}
\begin{align}
    &J_S^{B^+} =  \frac{\tau_{B^+}}{2^8 \pi^3 m_{B^+}^3} \bigg( \frac{m_{B^+}^2 - m_{K^+}^2}{m_b - m_s} \bigg)^2 \int dq^2~ q^2 \lambda^{1/2}(m_{B^+}^2,m_{K^+}^2,q^2)   |f_0^{B}(q^2)|^2, \notag \\
    &J_T^{B^+} = \frac{\tau_{B^+}}{3 \cdot 2^4 \pi^3 m_{B^+}^3 (m_{B^+} + m_{K^+})^2} \int dq^2~ q^2 \lambda^{3/2}(m_{B^+}^2,m_{K^+}^2,q^2)  |f_T^{B}(q^2)|^2.
\end{align}
where $\tau_B$ is the lifetime of the $B$ meson, $q^2$ is integrated over $0\le q^2\le(m_{B^+} - m_{K^+})^2$, and $\lambda(x,y,z) = x^2 + y^2 + z^2 - 2(xy + yz + zx)$.
The corresponding $J$ factors for $K^+ \to \pi^+ \nu \nu$ can be obtained by the replacements $B^+ \to K^+$, $K^+ \to \pi^+$, $b \to s$, and $s \to d$.

For the $K_L \to \pi^0 \nu \nu$ decay, we have~\cite{Buras:2024ewl} 
\begin{align}
    &J_S^{K_L} =  \frac{\tau_{K_L}}{2^9 \pi^3 m_{K^0}^3} \bigg( \frac{m_{K^0}^2 - m_{\pi^0}^2}{m_s - m_d} \bigg)^2 \int dq^2~ q^2 \lambda^{1/2}(m_{K^0}^2,m_{\pi^0}^2,q^2)   |f_0^{K}(q^2)|^2, \notag \\
    &J_T^{K_L} = \frac{\tau_{K_L}}{3 \cdot 2^5 \pi^3 m_{K^0}^3 (m_{K^0} + m_{\pi^0})^2} \int dq^2~ q^2 \lambda^{3/2}(m_{K^0}^2,m_{\pi^0}^2,q^2) |f_T^{K}(q^2)|^2,
\end{align}
where the $q^2$ integral is taken over the range $[0,(m_{K_L} - m_{\pi^0})^2]$.

\section{Thermally averaged cross sections \label{sec:thermal}}
Here we outline the calculation of $\gamma_i^{eq}$ for the processes in table~\ref{tab:process_LLQdH}.
By neglecting the masses, the squared amplitude for a process $i=\text{(A1)--(A10)}$, related to $C^{pr}_{\alpha \beta}$, is given by
\begin{align}
  \overline{\sum}|\mathcal{M}_i|^2 &= N_C N_L^2 \Big\{ |C_{\alpha \beta}^{pr}|^2 q_{12} q_{34} + |C_{\beta \alpha}^{pr}|^2 q_{14} q_{32} - \frac{1}{2} \mathrm{Re}[C_{\alpha \beta}^{pr} C_{\beta \alpha}^{pr *} ] (q_{12} q_{34} - q_{13} q_{24} + q_{14} q_{23}) \notag \\
  &- 2 \mathrm{Im}[C_{\alpha \beta}^{pr} C_{\beta \alpha}^{pr*} ] \epsilon^{\mu \nu \rho \sigma} q_{1 \mu} q_{2 \nu} q_{3 \rho} q_{4 \sigma} \Big\}.
\end{align}
We assign the external momenta $\{ q_1,q_2,q_3,q_4,q_5 \}$ to $\{ d_p, L_{\alpha}, Q_r, L_{\beta}, H \}$, respectively, and define $q_{ij}=(q_i+q_j)^2$.
Each $q_i$ may label either an initial or a final state, depending on the process, but $q_2$ is always assigned to an initial state.
The summation $\overline{\sum}$ is over the color ($N_C=3$), $\mathrm{SU}(2)_L$ ($N_L=2$), and spin degrees of freedom.
The CP-odd term, including the anti-symmetric tensor $\epsilon^{\mu \nu \rho \sigma}$, vanishes when we integrate over the full momentum space.

The phase space integral can be written by the Lorentz invariants for 2-to-3 scattering,
\begin{align}
 &s = (p_1+p_2)^2,    &&t_1 = (p_1-k_1)^2, && t_2 = (p_2-k_2)^2, && s_1 = (k_2 + k_3)^2,\notag\\
 &s_2 = (k_1 + k_3)^2;&&s_3 = (k_1 + k_2)^2, &&u_1 = (p_1 - k_2)^2, && u_2 = (p_2 - k_1)^2,
\end{align}
where $p_i$ ($k_i$) are the initial (final) momenta and five among them ({, the first five) are independent.
We then obtain
\begin{align}
  \gamma_{(A1)}^{eq} &= \gamma_{(A8)}^{eq} = \sum_{pr} \bigg\{ A_{\alpha \beta}^{pr} \int_{(2,3)} s s_3 + (A_{\beta \alpha}^{pr} + B_{\alpha \beta}^{pr}) \int_{(2,3)} t_1 t_2 \bigg\}, \notag \\
  \gamma_{(A2)}^{eq} &= \gamma_{(A6)}^{eq} = \sum_{pr} \bigg\{ B_{\alpha \beta}^{pr} \int_{(2,3)} s s_3 +(A_{\alpha \beta}^{pr} + A_{\beta \alpha}^{pr}) \int_{(2,3)} t_1 t_2 \bigg\}, \notag \\
  \gamma_{(A3)}^{eq} &= \gamma_{(A5)}^{eq} = \sum_{pr} \bigg\{ A_{\beta \alpha}^{pr} \int_{(2,3)} s s_3 +(A_{\alpha \beta}^{pr} + B_{\alpha \beta}^{pr}) \int_{(2,3)} t_1 t_2 \bigg\} , \notag \\
  \gamma_{(A4)}^{eq} &= \gamma_{(A7)}^{eq} = \gamma_{(A9)}^{eq} = \gamma_{(A10)}^{eq} = \sum_{pr} \bigg\{ (A_{\alpha \beta}^{pr} + A_{\beta \alpha}^{pr} + B_{\alpha \beta}^{pr}) \int_{(2,3)} (t_1 - t_2 + s_2) s_3 \bigg\},
\end{align}
where
\begin{align}
  A_{\alpha \beta}^{pr} = N_C N_L^2 \Big( |C_{\alpha \beta}^{pr}|^2 - \frac{1}{2} \mathrm{Re}[ C_{\alpha \beta}^{pr} C_{\beta \alpha}^{pr*} ] \Big), ~~~ B_{\alpha \beta}^{pr} = \frac{N_C N_L^2}{2} \mathrm{Re}[ C_{\alpha \beta}^{pr} C_{\beta \alpha}^{pr*} ] .
\end{align}
We have used the shorthand notations for $n \to m$ scatterings as 
\begin{align}
  \int_{(n,m)} \equiv \bigg[ \int [d \Pi]_{\mathrm{ini}} [f^{eq}]_{\mathrm{ini}} \bigg] \bigg[\int [d \Pi]_{\mathrm{fin}} (2 \pi)^4 \delta^4 \Big(\sum_{\mathrm{ini}} (p) - \sum_{\mathrm{fin}} (p) \Big) \bigg].
\end{align}
Since the initial and final momentum integrals are evaluated as 
\begin{align}
  &\int_{(2,3)} s s_3 = \frac{3}{4} \frac{T^{10}}{\pi^7}, ~~~\int_{(2,3)} t_1 t_2 = \frac{9}{32} \frac{T^{10}}{\pi^7}, ~~~\int_{(2,3)} (t_1-t_2+s_2)s_3 = \frac{3}{16} \frac{T^{10}}{\pi^7}, 
\end{align}
we consequently have
\begin{align}
  \sum_{i=1}^{10} \gamma^{eq}_{(Ai)} = \frac{81}{2} \frac{T^{10}}{\pi^7} \sum_{pr} \Big( |C_{\alpha  \beta }^{pr}|^2 + |C_{\beta  \alpha }^{pr}|^2 -\frac{1}{2} \mathrm{Re}[C_{\alpha  \beta }^{pr} C_{\beta  \alpha }^{pr *}] \Big).
\end{align}

\section{Loop functions \label{sec:twoloop}}

To derive the formulae of the neutrino masses in section~\ref{sec:neutrinomass}, we define the following integrals:
\begin{align}
  &I_1^{\mu \nu}(x,y,z,w) = \int \tilde{dl_1} \tilde{dl_2} \frac{l_1^\mu l_2^\nu}{l_2^2 - x} \frac{1}{(l_2-l_1)^2 - y} \frac{1}{l_1^2 - z} \frac{1}{l_1^2 - w}, \notag \\
  &I_2(x,y,z,w) = \int \tilde{dl_1} \tilde{dl_2} \frac{1}{l_2^2 - x} \frac{1}{(l_2-l_1)^2 - y} \frac{1}{l_1^2 - z} \frac{1}{l_1^2 - w}, \notag \\
  &I_3^{\mu \nu}(x,y,z,w) = \int \tilde{dl_1} \tilde{dl_2} \frac{l_1^\mu l_1^\nu}{l_2^2 - x} \frac{1}{(l_2-l_1)^2 - y} \frac{1}{l_1^2 - z} \frac{1}{l_1^2 - w}, \notag \\
  &I_4^{\mu \nu}(x,y,z,w) = \int \tilde{dl_1} \tilde{dl_2} \frac{l_2^\mu l_2^\nu}{l_2^2 - x} \frac{1}{(l_2-l_1)^2 - y} \frac{1}{l_1^2 - z} \frac{1}{l_1^2 - w},
\end{align}
where $\tilde{dl} = \frac{d^D l}{i(2\pi)^D} \mu^{2 \varepsilon}$.
The tensor functions can be reduced to scalar functions as 
\begin{align}
  I_i^{\mu \nu}(x,y,z,w) = g^{\mu \nu} I_i (x,y,z,w), ~~ (i=1,3,4).
\end{align}
After Wick rotation, we obtain
\begin{align}
  &DI_1(x,y,z,w) = \frac{1}{2(z-w)} \bigg[  J(y,z) - J(y,w) - J(x,z) + J(x,w) \notag \\
  &\qquad \qquad \qquad -(z-y+x) I(x,z,y) + (w-y+x)I(x,w,y) \bigg], \notag \\
  &I_2(x,y,z,w) = \frac{-1}{z-w} \bigg[ I(x,z,y) - I(x,w,y) \bigg], \notag \\
  &DI_3(x,y,z,w) = z I_2(x,y,z,w) - I(x,w,y), \notag \\
  &DI_4(x,y,z,w) = \frac{1}{z-w} \bigg[  J(y,z) - J(y,w)  -x \Big( I(x,z,y) - I(x,w,y) \Big) \bigg].
\end{align}
The master integrals are given by~\cite{Ford:1992pn,Martin:2001vx}
\begin{align}
  \kappa J(x) &= x (\overline{\ln}x-1), \notag \\
  \kappa^2 J(x,y) &= xy (\overline{\ln}x-1)(\overline{\ln}y-1), \notag \\
  \kappa^2 I(x,y,z) &= \frac{1}{2}(x-y-z)\overline{\ln}y \overline{\ln}z + \frac{1}{2}(y-x-z)\overline{\ln}x \overline{\ln}z + \frac{1}{2}(z-x-y)\overline{\ln}x \overline{\ln}y \notag \\
  &+ 2x \overline{\ln}x + 2y \overline{\ln}y + 2z \overline{\ln}z -\frac{5}{2}(x+y+z) - \frac{1}{2}\xi(x,y,z),
\end{align}
where $\kappa = 16 \pi^2$.
The Lobachevskiy's function $\xi(x,y,z)$ is given by
\begin{align}
  \xi(x,y,z) &= R \Big( 2 \ln\frac{z+x-y-R}{2z}\ln\frac{z+y-x-R}{2z} - \ln\frac{x}{z}\ln\frac{y}{z} - 2\mathrm{Li}_2 \frac{z+x-y-R}{2z} \notag \\
  &- 2\mathrm{Li}_2 \frac{z+y-x-R}{2z} + \frac{\pi^2}{3} \Big),
\end{align}
with 
\begin{align}
  R=\sqrt{x^2 + y^2 + z^2 - 2xy -2xz -2yz}.
\end{align}

The finite part of the Passarino--Veltmann functions~\cite{tHooft:1978jhc,Passarino:1978jh}, which are used in section~\ref{sec:0nuee}, are given by~\cite{Dreiner:2023yus}
\begin{align}
    &B_0 [p^2; m_a^2, m_b^2] = -\int^1_0 dx~\log \Big( \frac{p^2 x^2 - (p^2 + m_a^2 - m_b^2 ) x + m_a^2 - i \epsilon}{Q^2} \Big), \notag \\
    &B_1 [p^2; m_a^2, m_b^2] = \int^1_0 dx~x \log \Big( \frac{p^2 x^2 - (p^2 + m_a^2 - m_b^2 ) x + m_a^2 - i \epsilon}{Q^2} \Big).
\end{align}

\bibliographystyle{utphys31mod}

\bibliography{biblio}

\end{document}